\documentclass[10pt,oneside,final]{IEEEtran}

\usepackage{amssymb}
\usepackage{amsmath}
\usepackage{graphicx}
\usepackage{color}
\usepackage{multicol}
\usepackage{subfigure}
\usepackage{bm}
\usepackage{latexsym}
\usepackage{stfloats}
\usepackage{float}
\usepackage{exscale}
\usepackage{relsize}
\usepackage{cite}
\usepackage{cases}
\usepackage{algorithm}
\usepackage{algpseudocode}
\usepackage{amsfonts}
\usepackage{amsmath}
\usepackage{graphicx}
\usepackage{color}
\usepackage{multicol}
\usepackage{amssymb}
\usepackage{subfigure}
\usepackage{bm}
\usepackage{latexsym}
\usepackage{stfloats}
\usepackage{float}
\usepackage{exscale}
\usepackage{relsize}
\usepackage{cite}
\usepackage{epsfig}
\usepackage{epstopdf}
\usepackage{breqn}
\usepackage{enumerate}
\usepackage{multirow}
\usepackage[utf8]{inputenc}
\usepackage{algorithmicx}
\usepackage{CJK}
\usepackage{indentfirst}
\usepackage{amsmath}
\usepackage{cases}
\usepackage{amssymb}
\makeatletter

\newcommand{\Rmnum}[1]{\expandafter\@slowromancap\romannumeral #1@}
\makeatother

\makeatletter
\renewcommand{\maketag@@@}[1]{\hbox{\m@th\normalsize\normalfont#1}}%
\makeatother

\begin{document}
\title{
Beamforming Optimization for Active RIS-Aided Multiuser Communications With Hardware Impairments}
\author{
Zhangjie Peng, Zhibo Zhang, Cunhua Pan,~\IEEEmembership{Senior Member,~IEEE}, Marco Di Renzo,~\IEEEmembership{Fellow,~IEEE},~\\Octavia A. Dobre,~\IEEEmembership{Fellow,~IEEE},
and Jiangzhou Wang,~\IEEEmembership{Fellow,~IEEE}

\thanks{The work of Zhangjie Peng was supported in part by the Natural Science Foundation of Shanghai under Grant 22ZR1445600,  and in part by the National Natural Science Foundation of China under Grant 61701307. The work of Cunhua Pan was supported in part by the National Natural Science Foundation of China under Grants 62201137, 62331023 and 62350710796, National Key Research And Development Plan 2023YFB2905100, the Fundamental Research Funds for the Central Universities under Grant 2242022k60001, and the Research Fund of National Mobile Communications Research Laboratory, Southeast University under Grant 2023A03. The work of Octavia. A. Dobre was supported in part by the Natural Sciences and Engineering Research Council of Canada (NSERC), through its Discovery program. The work of Jiangzhou Wang was supported in part by the National Natural Science Foundation of China under Grants 62350710796. {\textit{(Corresponding authors: Cunhua Pan and Zhibo Zhang.)}}
	
	Zhangjie Peng is with the College of Information, Mechanical and Electrical
	Engineering, Shanghai Normal University, Shanghai 200234, China, also
	with the National Mobile Communications Research Laboratory, Southeast
	University, Nanjing 210096, China, and also with the Shanghai Engineering
	Research Center of Intelligent Education and Bigdata, Shanghai Normal
	University, Shanghai 200234, China (e-mail: pengzhangjie@shnu.edu.cn).
	
	Zhibo Zhang is with the College of Information, Mechanical and Electrical Engineering, Shanghai Normal University, Shanghai 200234, China (e-mail: 1000497171@smail.shnu.edu.cn).  
	
Cunhua Pan is with the National Mobile Communications Research Laboratory, Southeast University, Nanjing 210096, China (E-mail:cpan@seu.edu.cn). 

Marco Di Renzo is with Université Paris-Saclay, CNRS, CentraleSupélec,
Laboratoire des Signaux et Systèmes, 91192 Gif-sur-Yvette, France (e-mail:
marco.di-renzo@universite-paris-saclay.fr). 

Octavia A. Dobre is with the Faculty of Engineering and Applied Science, Memorial University, St. John’s, NL A1C 5S7, Canada (e-mail:odobre@mun.ca). 

Jiangzhou Wang is with the School of Engineering, University of Kent, CT2 7NT Canterbury, U.K. (e-mail: j.z.wang@kent.ac.uk).
}}

\maketitle
\newtheorem{lemma}{$\textbf{Lemma}$}
\newtheorem{theorem}{Theorem}
\newtheorem{remark}{Remark}
\newtheorem{corollary}{Corollary}
\newtheorem{proposition}{Proposition}
\newtheorem{proof}{Proof}
\allowdisplaybreaks[4]
 
\begin{abstract}
In this paper, we consider an active reconfigurable intelligent surface (RIS) to assist the multiuser downlink transmission in the presence of practical hardware impairments (HWIs), including the HWIs at the transceivers and the phase noise at the active RIS. The active RIS is deployed to amplify the incident signals to alleviate the multiplicative fading effect, which is a limitation in the conventional passive RIS-aided wireless systems. We aim to maximize the sum rate through jointly designing the transmit beamforming at the base station (BS), the amplification factors and the phase shifts at the active RIS. 
To tackle this challenging optimization problem effectively, we decouple it into two tractable subproblems. Subsequently, each subproblem is transformed into a second order cone programming problem. The block coordinate descent framework is applied to tackle them, where the transmit beamforming and the reflection coefficients are alternately designed. In addition, another efficient algorithm is presented to reduce the computational complexity. Specifically, by exploiting the majorization-minimization approach, each subproblem is reformulated into a tractable surrogate problem, whose closed-form solutions are obtained by Lagrange dual decomposition approach and element-wise alternating sequential optimization method. Simulation results validate the effectiveness of our developed algorithms, and reveal that the HWIs significantly limit the system performance of active RIS-empowered wireless communications. Furthermore, the active RIS noticeably boosts the sum rate under the same total power budget, compared with the passive RIS. 


\end{abstract}

\section{Introduction}
The future sixth generation (6G) of communication systems are anticipated to fulfill the increasing demand for high-quality data transmission and reliable wireless connectivity \cite{8869705}. 
Recently, 
reconfigurable intelligent surface (RIS) has emerged as a prospective technology for 6G networks \cite{9475160}. The RIS makes the radio propagation environment intelligently reconfigurable, which introduces a notable communication paradigm \cite{9136592,9847080,8910627,9475160}.
In particular, comprising numerous programmable and passive reflecting elements coupled with low power electronics, the RIS can be deployed to establish a virtual line-of-sight link from the source to the destination \cite{9475160}, which is able to improve the coverage and capacity of wireless networks \cite{9424177,9140329,9048622}. 
These appealing characteristics have inspired extensive research works on RIS to improve the performance of traditional communications, including simultaneous wireless information and power transfer (SWIPT) \cite{9110849}, mobile edge computing (MEC) \cite{9133107}, cognitive radio 
\cite{9146170}, physical layer security  
\cite{9201173}, visible light communication \cite{9474926}, and unmanned aerial vehicle systems 
\cite{9367288} 
. 
It has been verified that, with the aid of RISs, the system performance can be improved in terms of various design objectives, 
e.g., weighted sum rate (WSR) maximization \cite{9090356}, 
weighted minimum rate maximization \cite{9318531}, 
energy efficiency maximization in \cite{8741198}, 
transmit power minimization 
\cite{8811733}\cite{9180053}, 
and outage probability minimization \cite{9205879}. 

Despite the fact that the aforementioned RIS (i.e., {\textit{nearly passive RIS}}) possesses appealing advantages in various aspects, each element reflects the incident signals without performing any signal processing tasks and without signal amplification.
The reflected signals, however, suffer from the multiplicative fading effect, resulting in a higher path loss for the reflecting link compared to that of the direct link \cite{9306896}. Specifically, the signals reflected by a nearly passive RIS are characterized by a cascaded channel, which consists of the source-RIS link and the RIS-destination link \cite{8879620}. As a consequence, a sufficient number of nearly passive elements is necessary to obtain a high-quality reflected link, 
which usually results in a large surface size. As a remedy, a novel type of the RIS, named {\textit{active RIS}}, has been proposed to mitigate the multiplicative fading effect \cite{zhang2021active,9377648,9734027}. Specifically, an active RIS is equipped with active reflection-type amplifiers \cite{zhang2021active}. Therefore, it has the ability to reflect signals like a nearly passive RIS, and to amplify signals with an additional power supply. 
The contributions \cite{zhang2021active,9377648,9734027,9690473,9530750,xu2021resource} revealed that the performance gain obtained by active RIS outperforms passive RIS under the same total power consumption in most cases. 
The comparison between active and passive RISs systems in terms of energy efficiency was studied in \cite{fotock2023energy} and \cite{9947328}.
In \cite{9652031}, an active RIS was deployed for improving the secrecy performance by optimizing the transmit beamforming and the reflection coefficients of the active RIS. In \cite{9878164}, the receive beamforming and the reflection coefficients at the active RIS were jointly designed for minimizing the maximum computational latency in the active RIS-empowered MEC systems. An active RIS-aided SWIPT system was explored in \cite{9849458} to maximize the WSR and weighted sum-power.

It is worth noting that the existing researches about active RIS-aided communications \cite{zhang2021active,9377648,9734027,9652031,9878164,9690473,fotock2023energy,9947328,9849458,9530750,xu2021resource} are based on the ideal assumption that both transceivers and active RIS are equipped with perfect hardware. However, in practical communication systems, the hardware impairment (HWI) at the transceivers (T-HWI) is inevitable due to hardware non-linearity and the quantization errors \cite{6891254}. Unlike the additive noise at the receiver, the T-HWI distorts transmitted and received signals. Although various compensation algorithms were proposed to mitigate the performance degradation caused by HWI \cite{9781613}, it cannot be completely eliminated \cite{6891254}. Furthermore, owing to the infeasibility of high-precision configuration of RIS phase shifts, the HWI at the RIS (RIS-HWI) is non-negligible, and it is usually modeled as RIS phase noise. Different studies have demonstrated the negative impacts of T-HWI and/or RIS-HWI in passive RIS-assisted communication systems \cite{9159653,9239335,9650619,9390410,9374557,9775986}. Specifically, the passive RIS-aided wireless powered Internet of Things network was investigated in \cite{9775986}, revealing that both T-HWI and RIS-HWI resulted in a performance degradation. 
Accordingly, it is also necessary to take into account both T-HWI and RIS-HWI in active RIS-assisted communication systems. An active RIS-aided device-to-device communication in the presence of RIS-HWI was investigated in \cite{9840889}, the closed-form and asymptotic expressions of the ergodic sum rate were derived. The work in \cite{10025392} considered an active RIS-assisted single-input-single-output system with RIS-HWI, the effect of which on system performance was analyzed. However, as far as we know, the performance loss caused by both T-HWI and RIS-HWI in active RIS-aided communications remains uninvestigated.

Against this background, we focus on an active RIS-empowered multiuser multiple-input single-output (MISO) communication system in the presence of both T-HWI and RIS-HWI, and we investigate the impact of hardware imperfections on the sum rate performance. The transmit beamforming of the base station (BS) and the reflection coefficients of the active RIS are jointly designed to maximize the sum rate of all users. Different from existing contributions on active RIS-empowered communication networks \cite{zhang2021active,9377648,9734027,9652031,9878164,9690473,fotock2023energy,9947328,9849458,9530750,xu2021resource} under the ideal assumption of perfect transceivers hardware and ideal RIS, the non-negligible HWIs at both transceivers and active RIS are taken into account. The problem is challenging to tackle owing to the presence of HWIs and the dynamic noise introduced by active RIS, which cannot be directly addressed with the existing methods. The main contributions of this paper are outlined as below:
\begin{itemize}
	\item To the best of our knowledge, this is the first attempt to investigate the active RIS-empowered multiuser communications by taking both T-HWI and RIS-HWI into consideration. Specifically, the transmit beamforming and the reflection coefficients are jointly designed for maximizing the sum rate. 
	Nevertheless, the formulated problem is challenging to tackle owing to the non-differentiable nature of the objective function and the strong coupling among the optimization variables. 
	\item To efficiently deal with this sophisticated problem, the original optimization problem is firstly reformulated into an equivalent form through taking advantage of the fractional programming (FP) criterion and quadratic transform. We decouple the original problem into two more tractable subproblems. To be specific, two groups of auxiliary variables are introduced to  transform the BS beamforming optimization problem and the active RIS reflection coefficients design problem into second order cone programming (SOCP) problems. Subsequently, the formulated problems are addressed efficiently through invoking the block coordinate descent (BCD) framework, where the optimization variables are alternately updated.
    \item To further reduce the computational complexity, the minorization-maximization (MM)-based approaches are proposed for deriving a closed-form solution for each subproblem. Based on the classical Lagrangian dual decomposition approach and the MM method, a closed-form solution is derived for the BS beamforming matrix. Then, by exploiting the MM method to simplify the objective function and the constraints, we develop a highly effective element-wise alternating sequential optimization (ASO) approach to achieve a high-quality solution for the reflection coefficients. The designed algorithms can be directly extended to scenarios without taking into account the impact of HWIs, and to the single-user scenarios.
	\item Extensive simulation results confirm the superiority of active RIS-empowered communications over nearly passive RIS communications in terms of sum rate under the same total power budget. Furthermore, it is indicated that the performance loss caused by HWIs is non-negligible, which unveils the significance of considering both T-HWI and RIS-HWI in active RIS-assisted communications. 
	Finally, the convergence, the effectiveness and the low complexity of the developed approaches are elucidated as well.
\end{itemize}

The remainder of this paper is organized as follows. We present the system model of an active RIS-empowered multiuser MISO communication system in the presence of HWIs, then the sum rate maximization problem is formulated in Section \ref{Section 2}, followed by decoupling the original problem into two more tractable subproblems in Section \ref{Section 3}, then the BCD framework based optimization approach is designed and serves as a benchmark. In Section \ref{Section 4}, low-complexity algorithms are developed. Simulation results are shown to confirm the effectiveness of the designed approaches, along with insightful
discussions and observations in Section \ref{Section 5}. Finally, Section \ref{Section 6} concludes the paper. 

\textit{Notations:~}Constant, vector, matrix are denoted in italic, bold lower case letter, and bold upper case letter, respectively. For a complex value $x$, ${\rm{Re}}\left\{x\right\}$, $\left|x\right|$ and $\angle x$ represent the real part, modulus and angle of $x$, respectively. $(\cdot)^*$, $(\cdot)^T$, $(\cdot)^H$, $(\cdot)^{-1}$ and $(\cdot)^{\dag}$ represent the conjugate, transpose, Hermitian transpose, inverse and pseudo-inverse operations, respectively. ${\rm{Tr}}(\cdot)$, $\left\|\cdot\right\|_F$ and ${\rm{vec}}\left(\cdot\right)$ stand for the trace, Frobenius norm and vectorization operator of the input matrix. ${\rm{diag}}(\cdot)$ stands for a diagonal matrix whose diagonal entries are the entries of the input vector. $\widetilde {\rm{diag}}(\cdot)$ represents a diagonal matrix whose diagonal entries are the same as that of the input matrix. For two matrices $\bf{A}$ and $\bf{B}$, ${\bf{A}} \otimes{\bf{B}}$ and ${\bf{A}} \odot{\bf{B}}$ are the Kronecker product and Hadamard product, respectively. ${\nabla}{{ f _{\bf{a}}}}({{{{\bf{a}}}}})$ stands for the gradient of the function $f$ w.r.t. $\bf{a}$. 

\section{System Model}\label{Section 2}



\subsection{System Model}
Consider an active RIS-assisted multiuser MISO system, in which an $N$-antenna BS serves $K$ single-antenna users. In this system, an $M$-element active RIS is deployed to enhance the system performance. In particular, to strengthen the reflected signals, each RIS element is integrated with an active reflection-type amplifier supported by the power supply \cite{zhang2021active}. The phase shift matrix and the amplification matrix of the active RIS are denoted by ${\bf{\Theta }} \buildrel \Delta \over = {\rm{diag}}( {e^{j{\varphi _1}}}, \cdots , {e^{j{\varphi _M}}} )\in {{\mathbb C} ^{M \times M}}$ and ${\bf{A}} \buildrel \Delta \over = {\rm{diag}}\left( {{a_1}, \cdots ,{a_M}} \right) \in {{\mathbb R} _ + ^{M \times M}}$, respectively.

\begin{figure}[t] 
	\begin{center}
		\includegraphics[scale=0.55]{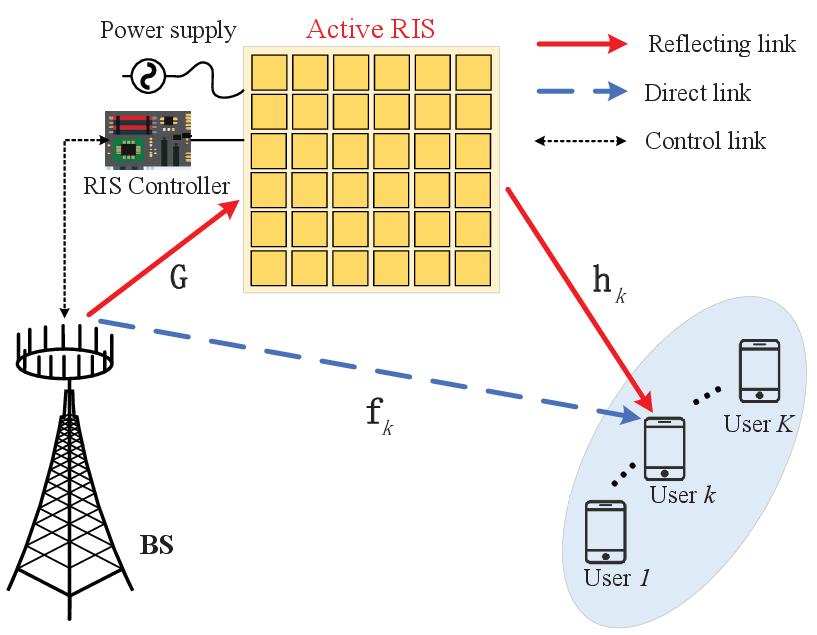}
		\caption{An illustration of an active RIS-assisted multiuser MISO communication system in the presence of transceiver HWIs.}
		\label{system_model}
	\end{center}
\end{figure}

The channel from the BS to the active RIS, the channel for the active RIS to the user $k$, and the direct channel from the BS to the user $k$, are modeled by ${{\bf{G}}} \in {{\mathbb C}^{M \times N}}$, ${{\bf{h}}_k} \in {{\mathbb C}^{M \times 1}}$, ${{\bf{f}}_k} \in {{\mathbb C}^{N \times 1}}$, respectively, as shown in Fig. \ref{system_model}. Based on the channel estimation techniques in \cite{9400843,9366805,8879620,10188853}, it is assumed that the channel state information (CSI) of all channels is perfectly known at the BS.
\subsection{HWIs Model}
In practice, due to the inherent imperfections of hardware, both the transmitted and the received signals are inevitably subject to the effect of the HWIs, which universally
exists in the practical communication systems. In the considered active RIS-aided communication system, the HWIs appear at both transceivers and active RIS. Therefore, there are two different types of HWIs, i.e., T-HWI and RIS-HWI.

For the T-HWI, it is modeled as an independent Gaussian random variable, which causes the mismatch between the received signal and the expected signal, or creates a distortion on the received signal during the reception processing \cite{9390410}. The power of the distortion noise is proportional to the signal power \cite{6891254}.

For the RIS-HWI, it can be modeled as a random diagonal phase noise matrix, which is denoted as ${\bf{\Phi }} \triangleq {\rm{diag}}\left( {\bm {\phi}}  \right)$ with $ {\bm {\phi}} = \left[ { {{\phi _1}, \ldots, {\phi _M}} } \right]^T$ and ${\phi _m} = {e^{j{\vartheta _m}}}$. ${\vartheta _m}$ is the phase noise at $m$-th element of the active RIS, and it is assumed to be uniformly distributed within $[-\pi/2, \pi/2]$ \cite{9390410}.
\subsection{Signal Transmission Model}

Different from the existing works on active RIS-assisted communications, such as \cite{zhang2021active,9377648,9734027,9652031,9878164,9690473,fotock2023energy,9947328,9849458,9530750,xu2021resource}, we consider the non-negligible HWIs at both transceivers and active RIS. More specifically, the
transmit signal of the BS is
\begin{equation}\label{BS_Signal}
	{{\bf{x}}} = {\widehat {\bf{x}}} + {\bm{\eta}_t}, \ \ \
	{\widehat {\bf{x}}} = \sum\limits_{k = 1}^K {{{\bf{w}}_k}{s_{k}}},
\end{equation}
where ${\bf{w}}_k\in {{\mathbb C}^{N \times 1}}$ stands for the beamforming vector adopted by the BS for user $k$,  $s_k$ denotes the corresponding data symbol with unit power and ${\bm{\eta}_t \in {{\mathbb C}^{N \times 1}}}$ is the transmit distortion noise, which is independent of $s_k$. The variance of each entry in ${\bm{\eta}_t}$ is proportional to the transmit power at the corresponding antenna, i.e., $\bm{\eta} _t \sim \mathcal{CN} \left( {\bm{0},\bm{\Upsilon} _t} \right)$, and $\bm{\Upsilon} _t$ is expressed as
\begin{equation}\label{7}
	\bm{\Upsilon} _t = 
	\kappa _t\widetilde {\rm{diag}} \left(\sum\limits_{k = 1}^{{K}} {{ {{{\bf{w}}_k}}{{{\bf{w}}_{k}^H}}}}\right),
\end{equation}
where $\kappa _t \in (0,1)$ is a scale factor which represents the severity of transmitter HWI at the BS.

Then, the signal reflected and amplified by the active RIS is expressed as
\begin{equation}
	{{\bf{y}}_{\rm RIS}} = \sum\limits_{k = 1}^K {{\bf{A\Theta \Phi G}}{{\bf{w}}_k}{s_k} \!+\! {\bf{A\Theta \Phi G}}{{\bm\eta} _t} \!+\! {\bf{A\Theta \Phi m}} + \bf n},\label{y_RIS}
\end{equation}
where the matrix ${\bf{A}} \buildrel \Delta \over = {\rm{diag}}\left( {{a_1}, \cdots ,{a_M}} \right) \in {{\mathbb R} _ + ^{M \times M}}$ stands for the amplification matrix, the matrix ${\bf{\Theta }} \buildrel \Delta \over = {\rm{diag}}( {e^{j{\varphi _1}}}, \cdots , {e^{j{\varphi _M}}} )\in {{\mathbb C} ^{M \times M}}$ represents the phase shift matrix of the active RIS. The vector $\bf{m}$ represents the thermal noise at the active RIS, it cannot be neglected and is modeled as an additive white Gaussian noise (AWGN), whose distribution is ${\bf{m}} \sim {\mathcal{CN}}({\bf{0}}_M,\sigma _{d}^2{{\bf{I}}_M})$. The vector $\bf n$ denotes the static noise and it is generally negligible compared to the dynamic noise $\bf A\Theta \Phi m$ \cite{6047578} .\footnote{The noise at the active RIS can be categorized into dynamic noise $\bf A\Theta \Phi m$ and static noise $\bf n$. In particular, the dynamic noise is introduced and amplified by the amplifiers at the active RIS, while the static noise is generated by the patch and the phase-shift circuit. The power of the static noise is too small and it is usually negligible in comparison to the dynamic noise $\bf A\Theta \Phi m$ \cite{6047578}. In addition, $\bf m$ is related to the input noise and the inherent device noise of the active RIS, and it is amplified by the active RIS. Furthermore, it has been verified by the experimental results in \cite{zhang2021active} that the dynamic noise is also amplified.} Thus, the signal power at the active RIS can be obtained as
\begin{align}
	{P_{{{\bf{y}}_{\rm RIS}}}} =& \sum\limits_{k = 1}^K {{\left\| {{\bf{A\Theta \Phi G}}{{\bf{w}}_k}} \right\|}^2} + \sigma _d^2{{\left\| {{\bf{A\Theta \Phi}}} \right\|}^2_F}  \nonumber
	\\
	&+{\rm{Tr}}\left({\bf{A\Theta \Phi G}}{\bm{\Upsilon} _t}{{\bf{G}}^H}{{\bf{\Phi }}^H}{{\bf{\Theta }}^H}{{\bf{A}}^H}\right).\label{power_RIS}
\end{align}

Furthermore, the signal power at the active RIS should not be larger than the amplification power budget $P_A$, i.e., ${P_{{{\bf{y}}_{\rm RIS}}}} \le P_A$. Specifically, the amplification power constraint is given by 
\begin{align}
	&{\rm{Tr}}\left({\kappa _t}{\bf{A\Theta \Phi G}}\left( {{\widetilde {\rm{diag}}}\left( { {{{\bf{W}}}{\bf{W}}^H} } \right)} \right){{\bf{G}}^H}{{\bf{\Phi }}^H}{{\bf{\Theta }}^H}{{\bf{A}}^H}\right)~~~~~~~\nonumber\\
	 &~~~~~~~~~~~~~~~~~+{{\left\| {{\bf{A\Theta \Phi G}}{{\bf{W}}}} \right\|}^2_F} + \sigma _d^2{{\left\| {{\bf{A\Theta \Phi}}} \right\|}^2_F} \le {P_A}.\label{RIS_power_constraint} 
\end{align}


Furthermore, the received signal at user $k$ is
\begin{align}\label{Desired_signal}
	{y_k} =& \underbrace {{\bf{g}}_k^H{{\bf{w}}_k}{s_k}}_{{\text{Desired~ signal}}} + \underbrace {{\bf{g}}_k^H{{\bm{\eta }}_t}}_{{\text{Transmitter~HWI}}} + \underbrace {{\bf{g}}_k^H\sum\limits_{ i= 1,
		i\ne k\hfill}^K {{{\bf{w}}_i}{s_i}} }_{{\text{Multiuser~ interference}}}\nonumber\\
	&+ \underbrace {{\bf{h}}_k^H{\bf{A\Theta \Phi m}}}_{{\text{Dynamic~noise}}} + \underbrace {{n_k}}_{{\text{Noise ~at~user}}~k}+ \underbrace {{\eta _{r,k}}}_{{\text{Receiver~HWI}}},
\end{align}
where ${\bf{g}}_k^H \triangleq \left( {{\bf{h}}_k^H{\bf{A\Theta \Phi G}} + {\bf{f}}_k^H} \right) \in {{\mathbb C}^{1 \times N}} $, 
${n_k}$ stands for the AWGN at user $k$ with distribution of $\mathcal{CN}(0,\sigma _{k}^2)$. We denote ${y_k} \triangleq \widetilde {{y}}_k + {\eta _{r,k}}$, where ${\eta _{r,k}}$ denotes the receive distortion noise at user $k$, which is independent of ${\widetilde {{y}}}_k$. Note that the variance of ${\eta _{r,k}}$ is proportional to the power of received signal. Hence, we have ${\eta _{r,k}} \sim \mathcal{CN}(0,{\gamma _{r,k}})$. Here, ${\gamma _{r,k}}$ is given by ${\gamma _{r,k}} = \kappa _{r, k}\mathbb{E}\left\{ {{\left| \widetilde {y}_k \right|}^2} \right\}$, where $\kappa _{r, k} \in (0,1)$ is a scale factor which accounts for the severity of the receiver HWI at the user $k$.

Correspondingly, according to \eqref{Desired_signal}, the SINR at user $k$ can be obtained as Equation \eqref{SINR} at the top of this page.
\begin{figure*}[t]
\begin{align}\label{SINR}
	\!\!\!{\gamma _k} \!=\! \frac{{{{\left| {{\bf{g}}_k^H{{\bf{w}}_k}} \right|}^2}}}{{{\bf{g}}_k^H \!\left( {{\kappa _{r, k}}{{\bf{w}}_k}{\bf{w}}_k^H \!+\! \left( {1 \!+\! {\kappa _{r, k}}} \right){\kappa _t}\widetilde {\rm{diag}}\left( {\sum\limits_{k = 1}^K {{{\bf{w}}_k}{\bf{w}}_k^H} } \right)} \right){{\bf{g}}_k}\! + \! \left({1 \!+\! {\kappa _{r, k}}}\right)\!\! \left(\! \sum\limits_{\scriptstyle i = 1\hfill\atop
	\scriptstyle i \ne k\hfill}^K \!\!\! {{{\left| {{\bf{g}}_k^H{{\bf{w}}_i}} \right|}^2}}  \!+\! \sigma _d^2{{\left\| {{\bf{h}}_k^H{\bf{A\Theta \Phi}}} \right\|}^2} \!+ \!\sigma _k^2\right)}}
\end{align}
\hrulefill
\end{figure*}

Therefore, the data rate (bps/Hz) of user $k$ is expressed as
\begin{equation}\label{Rate_k}
	{R_k} = {\log }_2\left( {1 + {\gamma _k}} \right).
\end{equation}
\subsection{Total Power Consumption Model}
The total power budget\footnote{To establish the total power consumption model, two underlying assumptions need to be satisfied: $i)$ the transmit amplifiers of the BS and the reflection-type amplifiers of the active RIS operate in the linear region of their transfer functions without constraints on the incident signal power. Therefore, it is unnecessary to consider the saturation threshold of the amplifiers. $ii)$ the circuit power is assumed to be independent of the communication rate. Consequently, the hardware-dissipated power can be approximated by a constant power offset \cite{8741198}.} for the considered active RIS-empowered system is composed of the transmit power of the BS, the amplification power of the active RIS, the hardware static power of the BS and of the active RIS \cite{8741198,9947328}. To sum up, the total power budget of the active RIS-empowered communication system is given by 
\begin{align}
	{P_{tot}} = {\xi _T}{P_T} + {\xi _A}{P_A} + {P_{{\rm{BS}}}} +  {P_{{\rm{RIS}}}},\label{active}
\end{align}
where ${\xi _T} \triangleq \nu _T^{ - 1}$ and ${\xi _A}  \triangleq \nu _{\rm{A}}^{ - 1}$ with $\nu _{\rm{T}} \in (0,1]$ and $\nu _{\rm{A}}\in (0,1]$ being the efficiency of the transmit
power amplifiers and the efficiency of the active RIS amplifiers, respectively. $P_T$ represents the transmit power at the BS in the active RIS system. ${P_{{\rm{BS}}}}$  and ${P_{{\rm{RIS}}}}$ represent the hardware static power of the BS and of the active RIS, respectively. In particular, the static power of active RIS depends on the number of elements. Therefore, the static power dissipated at an active RIS with $M$ elements can be expressed as ${P_{{\rm{RIS}}}} = M\left( {{P_{{\rm{SW}}}} + {P_{{\rm{DC}}}}} \right)$, where ${P_{{\rm{SW}}}}$ and ${P_{{\rm{DC}}}}$ represent the control circuit power and the direct current (DC) biasing power consumed by each active RIS element, respectively \cite{9734027}.\footnote{As for the passive RIS system, the total power budget of it can be written as
	$P_{tot}^P = {\xi _T}P_T^P + {P_{{\rm{BS}}}} + M{P_{{\rm{SW}}}},$
where $P_T^P$ represents the transmit power at the BS in the passive RIS system. From the above total power budget of the passive RIS system and the total power budget of the active RIS system in \eqref{active}, it can be observed that we can guarantee ${P_{tot}} = {P^P_{tot}}$ by setting ${\xi _T}P_T^P = {\xi _T}P_T + {\xi _A}{P_A} + M {P_{{\rm{DC}}}}$.}

\subsection{Phase Noise Processing}
Due to the randomness of the phase noise matrix, it is difficult to obtain the exact expression for the date rate. Inspired by \cite{6816003,9390410,10065384}, we exploit the expectation to take over the randomness of $\bf \Phi$. Then, the average phase distortion level \cite{10056867} at the active RIS can be calculated to obtain the approximate average rate.

By adopting \cite[Lemma 1]{6816003}, we have
\begin{align}
	{R_k} \approx {\widetilde R_k}= {\log _2}\left( {1 + \widetilde {{\gamma}}} _k \right),
\end{align}
where ${\widetilde R}_k$ denotes the approximate average rate \cite{9390410}, and
\begin{align}
\widetilde {{\gamma}} _k =& \frac{{{\varpi _1}}}{{{\varpi _2} + {\varpi _3} - {\varpi _1}}},\\
{\varpi _1} \triangleq&  {\mathbf{w}}_k^H{\mathbb{E}_{\mathbf{\Phi }}}\left\{ {{{\mathbf{g}}_k}{\mathbf{g}}_k^H} \right\}{{\mathbf{w}}_k},\\
{\varpi _2} \triangleq& \left( {1 + {\kappa _{r,k}}} \right){\text{Tr}} \left(\left( {\sum\limits_{k = 1}^K {{{\mathbf{w}}_k}{\mathbf{w}}_k^H}  + {\kappa _t}\widetilde {{\text{diag}}}\left( {\sum\limits_{k = 1}^K {{{\mathbf{w}}_k}{\mathbf{w}}_k^H} } \right)} \right)\right.\nonumber\\
&\times {\mathbb{E}_{\mathbf{\Phi }}}\left\{ {{{\mathbf{g}}_k}{\mathbf{g}}_k^H} \right\}\Big) ,\\
{\varpi _3} \triangleq& {\mathbb{E}_{\mathbf{\Phi }}}\left\{ {\left( {1 + {\kappa _{r,k}}} \right)\left( {\sigma _d^2{{\left\| {{\mathbf{h}}_k^H{\mathbf{A\Theta \Phi }}} \right\|}^2} + \sigma _k^2} \right)} \right\}.
\end{align}

Furthermore, ${\mathbb{E}_{\bf{\Phi }}\left\{ {{{\bf{g}}_k}{\bf{g}}_k^H} \right\}}$ can be calculated as follows.
\begin{align}
		&\!\!\!{\mathbb{E}_{\bf{\Phi }}\left\{ {{{\bf{g}}_k}{\bf{g}}_k^H} \right\}}\nonumber \\ &= {{\bf{G}}^H}{{{\bf{\Theta }}^H}{{\bf{A}}^H}}{\rm{diag}}\left( {{{\bf{h}}_k}} \right){{\mathbb{E}}_{\bm \phi} }\left\{ {{\bm\phi} {{\bm\phi} ^H}} \right\}{\rm{diag}}\left( {{\bf{h}}_k^H} \right){\bf{A\Theta G}}
		\nonumber \\ &+ 2{\mathop{\rm Re}\nolimits} \left\{ {{{\bf{G}}^H}{{{\bf{\Theta }}^H}{{\bf{A}}^H}}{\rm{diag}}\left( {{{\bf{h}}_k}} \right){{\mathbb{E}}_{\bm \phi} }\left\{ {{{\bm\phi} ^*}} \right\}{\bf{f}}_k^H} \right\} + {{\bf{f}}_k}{\bf{f}}_k^H.\label{expectation_gk}
\end{align}

To further calculate ${\mathbb{E}_{\bf{\Phi }}\left\{ {{{\bf{g}}_k}{\bf{g}}_k^H} \right\}}$, we first need to obtain ${{\mathbb{E}}_{\bm \phi} }\left\{ {{{\bm\phi}{\bm\phi} ^H}} \right\}$ and ${{\mathbb{E}}_{\bm \phi} }\left\{ {{{\bm\phi} ^*}} \right\}$.
Denote $\delta _{\vartheta} = \vartheta _i -\vartheta _j, \forall i,j = 1,2,\cdots, M$.
Note that $\vartheta _i$ and $\vartheta _j$ are uniformly distributed within $[-\pi/2,\pi/2]$, their probability density function can be expressed as $f\left( \vartheta _i \right) =\frac{1}{\pi}$.
Thus, $\delta _{\vartheta}$ obeys triangular distribution on $[-\pi,\pi]$, and its probability density function can be expressed as \cite{9390410}
\begin{align}
	f\left( \delta _{\vartheta} \right) =\begin{cases}
		\frac{1}{\pi ^2}\delta _{\vartheta}+\frac{1}{\pi},&		\delta _{\vartheta}\in \left[ -\pi ,0 \right],\\
		-\frac{1}{\pi ^2}\delta _{\vartheta}+\frac{1}{\pi},&		\delta _{\vartheta}\in \left[ 0,\pi \right].\\
	\end{cases}
\end{align}

Therefore, we have $\mathbb{E}_{\delta _{\vartheta}}\left\{ e^{j\vartheta _i-j\vartheta _j} \right\} =\mathbb{E}_{\delta _{\vartheta}}\left\{ e^{j\delta _{\vartheta}} \right\} =\int_{-\pi}^{\pi}{f\left( \delta _{\vartheta} \right)}e^{j\delta _{\vartheta}}d\delta _{\vartheta}=\frac{4}{\pi ^2}$, and
$\mathbb{E}_{\bm{\phi }}\left\{ \bm{\phi }\bm{\phi }^H \right\}$ can be given by
{\small	
	\begin{align}
		\mathbb{E}_{\bm{\phi }}\left\{ \bm{\phi }\bm{\phi }^H \right\} &= \left( \begin{matrix}
			1&				\cdots&		\mathbb{E}_{\delta _{\vartheta}}\left\{ e^{j\vartheta _M-j\vartheta _1} \right\}\\
			\mathbb{E}_{\delta _{\vartheta}}\left\{ e^{j\vartheta _1-j\vartheta _2} \right\}&		\cdots&		\mathbb{E}_{\delta _{\vartheta}}\left\{ e^{j\vartheta _M-j\vartheta _2} \right\}\\
			\vdots&			\ddots&		\vdots\\
			\mathbb{E}_{\delta _{\vartheta}}\left\{ e^{j\vartheta _1-j\vartheta _M} \right\}&			\cdots&		1\\
		\end{matrix} \right)\nonumber\\&=\mathbf{I}_M+\mathbf{J}, \label{expectation_1}
\end{align}
where 
\begin{align}
	{\left[ {\bf{J}} \right]_{(i,j)}} = \left\{ {\begin{array}{*{20}{c}}
			{0,~~~~i = j,}\\
			{\frac{4}{{{\pi ^2}}},~~i \ne j.}
	\end{array}} \right.
\end{align}

Furthermore, to calculate ${{\mathbb{E}}_{\bm \phi} }\left\{ {{{\bm\phi} ^*}} \right\}$, we can obtain $\mathbb{E}_{\vartheta _i}\left\{ e^{-j\vartheta _i} \right\} =\int_{-\frac{\pi}{2}}^{\frac{\pi}{2}}{f\left( \vartheta _i \right)}\left( \cos \vartheta _i-j\sin \vartheta _i \right) d\vartheta _i=\frac{2}{\pi}$.
Then, we have
\begin{align}
	\mathbb{E}_{\bm{\phi }}\left\{ \bm{\phi }^* \right\} =\frac{2}{\pi}{\mathbf{1}}_M,\label{expectation_2}
\end{align}
where ${\mathbf{1}}_M$ stands for the unit column vector with all elements being one.}

By substituting \eqref{expectation_1} and \eqref{expectation_2} into \eqref{expectation_gk}, we have
\begin{align}
		\mathbb{E}_{\bf{\Phi }}&\left\{ {{{\bf{g}}_k}{\bf{g}}_k^H} \right\}\nonumber\\
		=& {{\bf{G}}^H}{{{\bf{\Theta }}^H}{{\bf{A}}^H}}{\rm{diag}}\left( {{{\bf{h}}_k}} \right)\left( {{{\bf{I}}_M} + {\bf{J}}} \right){\rm{diag}}\left( {{\bf{h}}_k^H} \right){\bf{A\Theta G}} \nonumber\\
		&+ 2{\mathop{\rm Re}\nolimits} \left\{ {{{\bf{G}}^H}{{{\bf{\Theta }}^H}{{\bf{A}}^H}}{\rm{diag}}\left( {{{\bf{h}}_k}} \right)\frac{2}{\pi }{{\bf{1}}_M}{\bf{f}}_k^H} \right\} + {{\bf{f}}_k}{\bf{f}}_k^H
		\nonumber\\
		=& {{\bf{G}}^H}{{{\bf{\Theta }}^H}{{\bf{A}}^H}}{\rm{diag}}\left( {{{\bf{h}}_k}} \right){\bf{D}}{{\bf{D}}^T}{\rm{diag}}\left( {{\bf{h}}_k^H} \right){\bf{A\Theta G}}
		\nonumber\\
		&+ \left( {\frac{2}{\pi }{{\bf{G}}^H}{{{\bf{\Theta }}^H}{{\bf{A}}^H}}{{\bf{h}}_k} + {{\bf{f}}_k}} \right)\left( {\frac{2}{\pi }{\bf{h}}_k^H{\bf{A\Theta G}} + {\bf{f}}_k^H} \right)
		\nonumber\\
		=& {\widehat {\bf{f}}_k}\widehat {\bf{f}}_k^H + {\widehat {\bf{G}}_k}\widehat {\bf{G}}_k^H
		= {\overline {\bf{G}} _k}\overline {\bf{G}} _k^H,
\end{align}
where ${\overline {\bf{G}} _k} \triangleq \left[ {{{\widehat {\bf{f}}}_k},{{\widehat {\bf{G}}}_k}} \right] \in \mathbb{C}^{N\times (1 + M)}$, $\widehat {\bf{f}}_k^H \triangleq\frac{2}{\pi }{\bf{h}}_k^H{\bf{A\Theta G}} + {\bf{f}}_k^H$, $\widehat {\bf{G}}_k^H \triangleq {{\bf{D}}^T}{\rm{diag}}\left( {{\bf{h}}_k^H} \right){\bf{A\Theta G}}$ and ${\bf{D}}{{\bf{D}}^T} \triangleq {\rm{diag}}\left( {\left( {1 - \frac{4}{{{\pi ^2}}}} \right){{\bf{I}}_M}} \right)$.
Therefore, the approximate average rate can be written as ${\widetilde R_k}= {\log _2}\left( {1 + \widetilde {{\gamma}}} _k \right)$, where
\begin{align}
	\widetilde {{\gamma}} _k =& \frac{\widetilde {{\varpi _1}}}{{\widetilde {{\varpi _2}} + \widetilde {{\varpi _3}}- \widetilde {{\varpi _1}}}},\\
	\widetilde {{\varpi _1}} \triangleq&{{\bf{w}}_k^H{{\overline {\bf{G}} }_k}\overline {\bf{G}} _k^H{{\bf{w}}_k}},\\
	\widetilde {{\varpi _2}} \triangleq&\left( {1 \!+\! {\kappa _{r,k}}} \right) {\rm{Tr}}
	\left({\left( {{\bf{W}}{{\bf{W}}^{\rm{H}}}\! +\! {\kappa _{\rm{t}}}\widetilde {{\rm{diag}}}\left( {{\bf{W}}{{\bf{W}}^{\rm{H}}}} \right)} \right){{\overline {\bf{G}} }_k}\overline {\bf{G}} _k^H} \right),\\
	\widetilde {{\varpi _3}} \triangleq& \left( {1 + {\kappa _{r,k}}} \right) \times \nonumber \\
	&\hspace{-0.8cm}\big( {\sigma _d^2{\rm{Tr}}\left( {{{{\bf{\Theta }}^H}{{\bf{A}}^H}}{\rm{diag}}\left( {{{\bf{h}}_k}} \right)\left( {{{\bf{I}}_M} + {\bf{J}}} \right){\rm{diag}}\left( {{\bf{h}}_k^H} \right){\bf{A\Theta }}} \right) + \sigma _k^2} \big).
\end{align}

\subsection{Problem Formulation}
In this work, we aim for jointly designing the beamforming matrix at the BS, the amplification matrix and the phase shift matrix at the active RIS, i.e., $\{{\bf{W}}, {\bf{A}}, {\bf{\Theta}}\}$, to maximize the sum rate. 
Different from the nearly passive RIS design in prior works, the amplifiers at the active RIS introduce amplification power constraints rather than the unit modulus constraints. Therefore, the optimization problem can be formulated as
\begin{subequations}\label{Problem_formulaton}
	\begin{align}
		\!\!\!\!\!\!\mathop {{\rm{max}}}\limits_{{{\bf{W}}},{\bf{A}},{\bf{\Theta }}}~~~&{ \sum\limits_{k = 1}^K {{\widetilde R_k}}} \label{a}\\ \label{b}
		{\rm{s.}}{\rm{t.}}~~~~~&{\rm{Tr}}\left( {{\bf{W}}{{\bf{W}}^H}} \right) \le {P_{T}},\\ \label{c}
		&
		{P_{{{\bf{y}}_{\rm RIS}}}}\le {P_A},
	\end{align}
\end{subequations}
where ${\bf{W}}\buildrel \Delta \over = [{\bf{w}}_{1},\dots,{\bf{w}}_{K}]\in {{\mathbb C}^{N \times K}}$. Obviously, Problem \eqref{Problem_formulaton} is non-convex thanks to the coupled optimization variables. Furthermore, compared with nearly passive RIS, active RIS also introduces additional dynamic noise term, which makes the objective function more difficult. Compared
to the corresponding scenario without HWI, the objective
function and the constraints of Problem \eqref{Problem_formulaton} become more sophisticated. Consequently, Problem \eqref{Problem_formulaton} cannot be readily addressed through the existing approaches. In what follows, we propose efficient algorithms to solve it.

\section{BCD-SOCP Algorithm} \label{Section 3}
In this section, to decouple the BS beamforming variables and the RIS reflection coefficient variables, Problem \eqref{Problem_formulaton} is firstly reformulated into a more tractable equivalent form with the aid of the FP method. Then, two subproblems are alternately
solved by exploiting the BCD framework.

\subsection{Problem Reformulation}
To solve the non-convex problem effectively, the original Problem \eqref{Problem_formulaton} is reformulated by means of the FP approach \cite{8314727}, where the optimization variables are decoupled. Notice that the amplification matrix ${\bf{A}}$ and the phase shift matrix ${\bf{\Theta}}$ always present in a product form, it can be expressed as ${\bf{\Psi }} = {\bf{A\Theta }} = {\rm{diag}}\left({{a_1}{e^{j{\varphi _1}}}, \cdots ,{a_M}{e^{j{\varphi _M}}}} \right) \in {{\mathbb{C}}^{M \times M}}$. Then, we introduce a lemma as below.

\begin{lemma}\label{lemma 001}
Upon introducing two groups of auxiliary variables ${\mathcal{U}} = \left\{ {{{\bf u}_k} \in {\mathcal{U}},\forall k \in {\mathcal{K}}} \right\}$ and ${\mathcal{V}} = \left\{ {{v_k} \in {\mathcal{V}},\forall k \in {\mathcal{K}}} \right\}$, and adopting the quadratic transform \cite{8314727}, Problem \eqref{Problem_formulaton} is equivalently reformulated as
	\begin{subequations}\label{Problem_reformulaton}
		\begin{align}
			\mathop {{\rm{max}}}\limits_{{{\bf{W}}, {\bf{\Psi }}, {\mathcal{U}}, \mathcal{V}}}~~~&{ \sum\limits_{k = 1}^K {r_k}\left( {{\bf{W}},{\bf{\Psi }},{\mathcal{U}},\mathcal{V}} \right)} \label{Reformulation_a}\\ \label{Reformulation_b}
			{\rm{s.}}{\rm{t.}}~~~~~&{\rm{Tr}}\left( {{\bf{W}}{{\bf{W}}^H}} \right) \le {P_{T}},\\ \label{Reformulation_c}
			&
			{P_{{{\bf{y}}_{\rm RIS}}}}\le {P_A},
		\end{align}
	\end{subequations}
where $\mathop {\max }\limits_{{\cal U},{\cal V}} \left\{ {{r_k}\left( {{\bf{W}},{\bf{\Psi }},{\cal U},{\cal V}} \right)} \right\} = \frac{1}{{{{\log }_2}e}}\widetilde {{R}}_k$ holds, and
\begin{align}
	\!\!\!\!{r_k}\left( {{\bf{W}},{\bf{\Psi }},{\mathcal{U}},\mathcal{V}} \right) =& \log \left( {1 + {v_k}} \right) \!-\! {v_k}\nonumber \\ 
	&\hspace{-2.8cm}+ 2\sqrt {\left( {1 + {v_k}} \right)} {\mathop{\rm Re}\nolimits} \left\{ {{\bf{u}}_k^H\overline {\bf{G}} _k^H{{\bf{w}}_k}} \right\}\!-\!{\bf{u}}_k^H{{\bf{u}}_k}{\widetilde {{\varpi _2}}}
	-{\bf{u}}_k^H{{\bf{u}}_k}{\widetilde {{\varpi _3}}},\label{r_k}\\
	{\bf u}_k^{\star} =&~  \frac{{\sqrt {\left( {1 + {v_k}} \right)}~ \overline {\bf{G}} _k^H{{\bf{w}}_k}}}{{\widetilde {{\varpi _2}} + \widetilde {{\varpi _3}}}} \in {\mathbb{C} }^{\left( {M + 1} \right) \times 1}, \label{optimal_u}\\
	v_k^{\star} =&~ \frac{\widetilde {{\varpi _1}}}{{\widetilde {{\varpi _2}} + \widetilde {{\varpi _3}}- \widetilde {{\varpi _1}}}}.\label{optimal_v}
\end{align}

\end{lemma}


Compared with the original objective function of Problem \eqref{Problem_formulaton}, the objective function of Problem \eqref{Problem_reformulaton} is more tractable, although more optimization variables are introduced. To deal with Problem \eqref{Problem_reformulaton}, we develop the BCD method to maximize the objective function in Problem \eqref{Problem_formulaton} by alternately designing each group of variables, while the other groups of variables keep fixed. Since the optimal ${{\bf u}_k}$ and optimal ${v_k}$ in each iteration can be obtained in \eqref{optimal_u} and \eqref{optimal_v}, respectively, the remaining task is to design the BS beamforming matrix $\bf{W}$ as well as the RIS reflection coefficient matrix ${\bf{\Psi }}$. Let us focus on optimizing ${\bf{W }}$ and ${\bf{\Psi }}$ in the following.

\subsection{Optimize the BS Beamforming Matrix $\bf{W}$} \label{BCD-SOCP-W}
In this subsection, the BS beamforming matrix $\bf{W}$ is optimized while keeping ${\bf{\Psi }}, ~{\mathcal{U}}$ and $\mathcal{V}$ fixed. Denote ${\bf{w}}\buildrel \Delta \over ={{\rm {vec}}{(\bf W)}}$ and introduce a selection vector $t_k \in {\mathbb{R}}^{K \times1}$, where all elements are zero except for the $k$-th element, which is one. Then, Lemma \ref{Lemma 2} is introduced.
\begin{lemma}\label{Lemma 2}
For a fixed RIS reflection coefficient matrix ${\bf{\Psi }}$ and auxiliary variables ${\mathcal{U}}$ and ${\mathcal{V}}$, the subproblem for the optimization of the BS beamforming vector $\bf{w}$ is transformed into an equivalent problem as follows:
\begin{subequations}\label{Problem_reformulaton_BS2}
	\begin{align}
		\mathop {{\rm{min}}}\limits_{{{\bf{w}}}}~~~&{{{\bf{w}}^H}{\bf{\tilde \Xi w}}  -2{\rm{Re}}\left\{ {{{\bm{\omega }}^H}{\bf{w}}} \right\} - c} \label{Reformulation_BS2_a}\\ \label{Reformulation_BS2_b}
		{\rm{s.}}{\rm{t.}}~~~& {{\bf{w}}^H{{\bf{w}}}} \le {P_{T}},\\ \label{Reformulation_BS2_c}
		& {{{\bf{w}}^H}{\bf{\Gamma w}}}  \le {P_m},
	\end{align}
\end{subequations}
\vspace{-0.2cm}
where 
\begin{align}
	{\bm{\omega }} \triangleq& \sum\limits_{k = 1}^K {{\text{vec}}\left( {{\mathbf{\Omega }}_k^H} \right)}  \in {\mathbb{C}^{NK \times 1}}, \\ 
	\widetilde {\mathbf{\Xi }} \triangleq& \sum\limits_{k = 1}^K {\left( {{{\mathbf{I}}_K} \otimes {{\mathbf{\Xi }}_k}} \right)} \in {{\mathbb{C}}^{NK\times NK}},\\
	{{\mathbf{\Omega }}_k} \triangleq& \sqrt {\left( {1 + {v_k}} \right)} {{\mathbf{t}}_k}{\mathbf{u}}_k^H\overline {\mathbf{G}} _k^H,\\
	{{\mathbf{\Xi }}_k} \triangleq& \left( {1 + {\kappa _{{\text{r,}}k}}} \right){\mathbf{u}}_k^H{{\mathbf{u}}_k}\left( {{{\overline {\mathbf{G}} }_k}\overline {\mathbf{G}} _k^H + {\kappa _{\text{t}}}\widetilde {{\rm{diag}}}\left( {{{\overline {\mathbf{G}} }_k}\overline {\mathbf{G}} _k^H} \right)} \right),\\
	{c_k} \triangleq& \sum\limits_{k = 1}^K\big( {\log \left( {1 + {v_k}} \right) - {v_k} - \left( {1 + {\kappa _{r,k}}} \right){\mathbf{u}}_k^H{{\mathbf{u}}_k}\times } \nonumber \\  
	&\hspace{-0.8cm} {\left( {\sigma _d^2{\text{Tr}}\left( {{{\bf{\Psi }}^H}{\rm{diag}}\left( {{{\bf{h}}_k}} \right)\!\left( {{{\mathbf{I}}_M} + {\mathbf{J}}} \right){\rm{diag}}\left( {{\mathbf{h}}_k^H} \right){\mathbf{\Psi }}} \right) \!+\! \sigma _k^2} \right)} \big),\\
	{\bf{\Gamma }} \triangleq& {{\bf{I}}_K} \otimes \left({{\bf{G}}^H}{{\bf{\Psi }}^H}{\bf{\Psi G}} + {\kappa _t}{\widetilde{\rm{diag}}}\left( {{{\bf{G}}^H}{{\bf{\Psi }}^H}{\bf{\Psi G}}} \right)\right),\\
	{P_m} \triangleq&{P_A} - \sigma _d^2{\left\| {\bf{\Psi }} \right\|^2_F}.
\end{align}

\end{lemma}

The reformulation in \eqref{Problem_reformulaton_BS2} is an SOCP problem, which can be tackled through exploiting the standard optimization packages, such as CVX \cite{CVX}.

\vspace{-0.3cm}
\subsection{Optimize the RIS Reflection Coefficient Matrix ${\bf{\Psi }}$} \label{SOCP phi}
In this subsection, the RIS reflection coefficient matrix $\bf{\Psi}$ is optimized while ${\bf{W}},~{\mathcal{U}}$ and $\mathcal{V}$ are kept fixed. Define ${\bm{\psi}}=\left[ {{a_1}{e^{j{\varphi _1}}}, \cdots ,{a_M}{e^{j{\varphi _M}}}} \right]^T \in {\mathbb{C}}^{M\times 1}$ as the diagonal entries of ${\bf{\Psi}}$, i.e., ${\bf{\Psi}}={\rm{diag}}({\bm{\psi}})$. Then, we introduce a lemma as below.

\begin{lemma}
For a fixed BS beamforming matrix ${\bf{W }}$ and auxiliary variables ${\mathcal{U}}$ and ${\mathcal{V}}$, the subproblem for the optimization of active RIS reflection coefficient vector ${\bm{\psi}}$ is equivalently transformed into the following form:
\begin{subequations}\label{Problem_reformulaton_RIS2}
	\begin{align}
		\mathop {{\rm{min}}}\limits_{{{\bm{\psi}}}}~~~&{{\bm{\psi }}^H}{\bf{\Delta}}{\bm {\psi }} - 2{\mathop{\rm Re}\nolimits} \left\{ {{\bm{\psi }}^H{{\bm{\alpha }}}} \right\} - d \label{Reformulation_RIS_a2}\\ \label{Reformulation_RIS_b2}
		{\rm{s.}}{\rm{t.}}~~~&{{\bm{\psi}} ^H}{\bf{\Lambda }}{\bm{\psi}}  \le {P_A},
	\end{align}
\end{subequations}
where 
\begin{align}
	{\mathbf{\Delta }} \triangleq& \sum\limits_{k = 1}^K {\left( {{{\mathbf{\Delta }}_{1,k}} + {{\mathbf{\Delta }}_{2,k}}} \right)},\\
	{{\bm{\alpha }}^H} \triangleq& \sum\limits_{k = 1}^K {\left( {\sqrt {\left( {1 + {v_k}} \right)} {\bm{\alpha }}_{1,{\bm \psi} ,k}^H - {\bm{\alpha }}_{2,{\bm \psi} ,k}^H} \right)}\\ 
	{{\mathbf{Q}}_k} \triangleq& {\mathbf{u}}_k^H{{\mathbf{u}}_k}\left( {1 + {\kappa _{r,k}}} \right)\left( {{\mathbf{W}}{{\mathbf{W}}^H} + {\kappa _t}\widetilde {{\rm{diag}}}\left( {{\mathbf{W}}{{\mathbf{W}}^H}} \right)} \right),\\
	{{\mathbf{u}}_k} \triangleq& {\left[ {\begin{array}{*{20}{c}}
	{{u_{\psi ,k}}}&{{{\mathbf{u}}^T_{\psi ,k}}} 
		\end{array}} \right]^T},\\
	{{\mathbf{\Delta }}_{1,k}} \triangleq& \left( {\frac{4}{{{\pi ^2}}}{{\mathbf{h}}_k}{\mathbf{h}}_k^H} \right) \odot {\left( {{\mathbf{G}}{{\mathbf{Q}}_k}{{\mathbf{G}}^H}} \right)^T} \nonumber\\
	+& \left( {{\text{diag}}\left( {{{\mathbf{h}}_k}} \right){\mathbf{D}}{{\mathbf{D}}^T}{\text{diag}}\left( {{\mathbf{h}}_k^H} \right)} \right) \odot {\left( {{\mathbf{G}}{{\mathbf{Q}}_k}{{\mathbf{G}}^H}} \right)^T},\\
	{{\mathbf{\Delta }}_{2,k}} \triangleq& \left( {\sigma _d^2\left( {1 + {\kappa _{r,k}}} \right){\mathbf{u}}_k^H{{\mathbf{u}}_k}{\rm{diag}}\left( {{{\mathbf{h}}_k}} \right)\left( {{{\mathbf{I}}_M} + {\mathbf{J}}} \right){\rm{diag}}\left( {{\mathbf{h}}_k^H} \right)} \right) \nonumber \\
	&\odot { {{{\mathbf{I}}_M}} },\\
	{\bm{\alpha }}_{1,{\bm \psi} ,k}^H \triangleq& \frac{2}{\pi }u_{{\bm \psi} ,k}^ * {\mathbf{h}}_k^H{\rm{diag}}({\mathbf{G}}{{\mathbf{w}}_k}) \nonumber\\
	&+ {\mathbf{u}}_{{\bm \psi} ,k}^H{{\mathbf{D}}^T}{\rm{diag}}\left( {{\mathbf{h}}_k^H} \right){\rm{diag}}({\mathbf{G}}{{\mathbf{w}}_k}),\\
	{\bm{\alpha }}_{2,{\bm \psi} ,k}^H \triangleq& \frac{2}{\pi }{\mathbf{h}}_k^H{\rm{diag}}\left( {{\mathbf{G}}{{\mathbf{Q}}_k}{{\mathbf{f}}_k}} \right),\\
	{d_{1,{\bm \psi} ,k}} \triangleq& u_{{\bm \psi} ,k}^ * {\mathbf{f}}_k^H{{\mathbf{w}}_k},~~{d_{2,{\bm \psi} ,k}} \triangleq {\mathbf{f}}_k^H{{\mathbf{Q}}_k}{{\mathbf{f}}_k},\\
	{d} \triangleq& \sum\limits_{k = 1}^K \left( \log \left( {1 + {v_k}} \right) - {v_k} - \left( {1 + {\kappa _{r,k}}} \right)\sigma _k^2{\mathbf{u}}_k^H{{\mathbf{u}}_k} \right.\nonumber\\
	&\left.+ 2\operatorname{Re} \left\{ {\sqrt {\left( {1 + {v_k}} \right)} {d_{1,\psi ,k}}} \right\} - {d_{2,\psi ,k}} \right),\\
	{\bf{\Lambda }} \triangleq& \left( {{\bf{G}}\left( {{\bf{W}}{{\bf{W}}^H} + {\kappa _t}{\widetilde{\rm{diag}}}\left( {{\bf{W}}{{\bf{W}}^H}} \right)} \right){{\bf{G}}^H} + \sigma _d^2{{\bf{I}}_M}} \right)\nonumber \\ &\odot {{\bf{I}}_M}.
\end{align}

\end{lemma}

The reformulation in \eqref{Problem_reformulaton_RIS2} is an SOCP problem, which can be addressed by exploiting standard optimization tools. 

In summary, Problem \eqref{Problem_reformulaton} can be tackled by Algorithm \ref{SOCP Algorithm}, the details of which are sketched as Algorithm \ref{SOCP Algorithm}.

\begin{algorithm}[t] 
	\caption{SOCP-based BCD Algorithm for Solving Probelm \eqref{Problem_reformulaton}} 
	\label{SOCP Algorithm} 
	
	\textbf{Initialize}: Initialize feasible ${\bf{w}}^{(0)}$ and ${\bm{\psi}}^{(0)}$, set the iteration number $n = 0$.
	\begin{algorithmic}[1]
		\Repeat 
		\State Given ${\bf{w}}^{(n)}$ and ${\bm{\psi}}^{(n)}$, evaluate the auxiliary variables ${\mathcal{U}}^{(n+1)}$ by \eqref{optimal_u} and 
		$\mathcal{V}^{(n+1)}$ by \eqref{optimal_v};
		\State Given ${\bm{\psi}}^{(n)}$, $\mathcal{U}^{(n+1)}$ and ${\mathcal{V}}^{(n+1)}$, obtain ${\bf{w}}^{(n+1)}$ by solving Problem \eqref{Problem_reformulaton_BS2} with CVX;
		\State Given ${\bf{w}}^{(n+1)}$, $\mathcal{U}^{(n+1)}$ and ${\mathcal{V}}^{(n+1)}$, obtain ${\bm{\psi}}^{(n+1)}$ by solving Problem \eqref{Problem_reformulaton_RIS2} with CVX;
		\State Set $n \gets n+1$;
		\Until The value of the objective function in \eqref{Problem_reformulaton} converges.
	\end{algorithmic}
\end{algorithm}

\section{A Low-Complexity Algorithm}\label{Section 4}
Note that solving the SOCP problem using CVX tools results in a large computational complexity. To deal with this issue, we develop a low-complexity MM-based approach with closed-form solutions to alternately obtain the BS transmit beamforming matrix and the active RIS reflection coefficient matrix by employing the BCD framework. Specifically, the convex subproblem of the BS beamforming matrix is solved by applying the MM-based Lagrangian dual decomposition method, and then, an effective MM-based element-wise ASO algorithm \cite{7946256} is proposed to achieve high-quality solutions for the active RIS reflection coefficient matrix.

\subsection{Optimize the BS Beamforming Matrix $\bf{W}$}\label{bisection search}
In this subsection, we present low-complexity algorithms to obtain nearly optimal closed-form solutions by using the Lagrangian dual decomposition approach \cite{boyd2004convex} along with the MM method. In particular, the subproblem in \eqref{Problem_reformulaton_BS2} is a convex SOCP problem, which can be tackled as described in Section \ref{BCD-SOCP-W}. Nevertheless, the computational burden of tackling the SOCP problem is heavy. To reduce the complexity, an MM-based algorithm with low complexity is developed to obtain a closed-form solution through using Lagrangian dual decomposition approach. To this end, a lemma is provided as below.

\begin{lemma}\label{lemma 1}
	Let ${f_{\bf{\Gamma }}}\left( {\bf{w }} \right) = {{\bf{w}}^H}{\bf{\Gamma w}}$, ${\bf{\Gamma }} \succeq \bf{0}$, ${{\bf{Z}}_{\bf{\Gamma }}} = {\lambda _{\bf{\Gamma }}}{\bf{I}}_{NK}$, where ${\lambda _{\bf{\Gamma }}}$ is the maximum eigenvalue of ${\bf{\Gamma }}$. Then, for any given solution ${{\bf{w}}^{\left( t \right)}}$ at the $t$-th iteration and for any feasible ${\bf{w}}$, there exists
	\begin{align}\label{lemma 1 equation}
		{{\tilde f}_{\bf{\Gamma }}}\left( {{\bf{w}}\left| {{{\bf{w}}^{\left( t \right)}}} \right.} \right) =& {{\bf{w}}^H}{{\bf{Z}}_{\bf{\Gamma }}}{\bf{w}} + {\left( {{{\bf{w}}^{\left( t \right)}}} \right)^H}\left( {{{\bf{Z}}_{\bf{\Gamma }}} - {\bf{\Gamma }}} \right){{\bf{w}}^{\left( t \right)}} \nonumber\\
		&- 2{\mathop{\rm Re}\nolimits} \left\{ {{{\left( {{{\bf{w}}^{\left( t \right)}}} \right)}^H}\left( {{{\bf{Z}}_{\bf{\Gamma }}} - {\bf{\Gamma }}} \right){\bf{w}}} \right\},
	\end{align}
which meets the following conditions:
\begin{enumerate}
	\item [1)] 
	${{\tilde f}_{\bf{\Gamma }}}\left( { {{{\bf{w}}^{\left( t \right)}}} \left| {{{\bf{w}}^{\left( t \right)}}} \right.} \right) = {f_{\bf{\Gamma }}}\left( {{{\bf{w}}^{\left( t \right)}}} \right)$;      
	\item [2)]
	${\nabla _{\bf{w}}}{{\tilde f}_{\bf{\Gamma }}}{\left( {{\bf{w}}\left| {{{\bf{w}}^{\left( t \right)}}} \right.} \right)_{{\bf{w}} = {{\bf{w}}^{\left( t \right)}}}} = {\nabla _{\bf{w}}}{f_{\bf{\Gamma }}}{\left( {{{\bf{w}}^{\left( t \right)}}} \right)_{{\bf{w}} = {{\bf{w}}^{\left( t \right)}}}}$;
	\item [3)]
	${{\tilde f}_{\bf{\Gamma }}}\left( {{\bf{w}}\left| {{{\bf{w}}^{\left( t \right)}}} \right.} \right) \ge {f_{\bf{\Gamma }}}\left( {\bf{w}} \right)$.
\end{enumerate}
\end{lemma}


According to Lemma \ref{lemma 1}, the constraint in \eqref{Reformulation_BS2_c} is replaced by
\begin{align}\label{replaced_Reformulation_BS2_c}
	{{\bf{w}}^H}{{\bf{Z}}_{\bf{\Gamma }}}{\bf{w}} - 2{\mathop{\rm Re}\nolimits} \left\{ {{{\left( {{{\bf{w}}^{\left( t \right)}}} \right)}^H}\left( {{{\bf{Z}}_{\bf{\Gamma }}} - {\bf{\Gamma }}} \right){\bf{w}}} \right\} \le {{\tilde P}_m},
\end{align}
where ${{\tilde P}_m} = {P_m} - {\left( {{{\bf{w}}^{\left( t \right)}}} \right)^H}\left( {{{\bf{Z}}_{\bf{\Gamma }}} - {\bf{\Gamma }}} \right){{\bf{w}}^{\left( t \right)}}$. Therefore, Problem \eqref{Problem_reformulaton_BS2} is transformed as follows
\begin{align} \label{replaced_Problem_reformulaton_BS2}
	\mathop {{\rm{min}}}\limits_{{{\bf{w}}}}~~~&{{{\bf{w}}^H}{\bf{\tilde \Xi w}}  -2{\rm{Re}}\left\{ {{{\bm{\omega }}^H}{\bf{w}}} \right\} - c}~~~\notag\\
	{\rm{s.}}{\rm{t.}}~~~
	&\eqref{Reformulation_BS2_b}, 
	\eqref{replaced_Reformulation_BS2_c}.
\end{align}

Problem \eqref{replaced_Problem_reformulaton_BS2} is a convex problem and the Slater’s condition is met, thus the dual gap between Problem
\eqref{replaced_Problem_reformulaton_BS2} and the corresponding dual problem is zero. Therefore, the optimal solution can be found by dealing with the corresponding dual problem more simply than the original problem. Suppose that the optimal solution of Problem \eqref{replaced_Problem_reformulaton_BS2} is ${\bf w}^\star$, according to whether the transmit power constraint is active at ${\bf w}^\star$ or not, two cases can occur.

$~\textit{1) Case \uppercase\expandafter{\romannumeral1}:}$ Assume that the BS transmit power constraint \eqref{Reformulation_BS2_b} is inactive at ${\bf w}^\star$, i.e., ${\bf w}^\star{{\bf w}^\star}^H < P_T$ holds. Problem \eqref{replaced_Problem_reformulaton_BS2} can be reformulated as the following equivalent form.
\begin{align} \label{replaced_Problem_reformulaton_BS2_case1}
	\mathop {{\rm{min}}}\limits_{{{\bf{w}}}}~~~&{{{\bf{w}}^H}{\bf{\tilde \Xi w}}  -2{\rm{Re}}\left\{ {{{\bm{\omega }}^H}{\bf{w}}} \right\} - c}\notag\\
	{\rm{s.}}{\rm{t.}}~~~
	&\eqref{replaced_Reformulation_BS2_c}.
\end{align}

Upon introducing the Lagrange multiplier $\mu$ associated with the constraint \eqref{replaced_Reformulation_BS2_c}, the Lagrangian function of Problem \eqref{replaced_Problem_reformulaton_BS2_case1} is expressed as
\begin{align}\label{Lagrangian function case 1}
	\!\!{\mathcal{L}}\left( {{\bf{w}},\mu } \right) =& {{\bf{w}}^H}\left( {{\bf{\tilde \Xi }} + \mu {{\bf{Z}}_{\bf{\Gamma }}}} \right){\bf{w}} - 2{\mathop{\rm Re}\nolimits} \left\{ {{{\bm{\omega }}^H}{\bf{w}}} \right\} \nonumber\\
	-& 2\mu {\mathop{\rm Re}\nolimits} \left\{ {{{\left( {{{\bf{w}}^{\left( t \right)}}} \right)}^H}\left( {{{\bf{Z}}_{\bf{\Gamma }}} - {\bf{\Gamma }}} \right){\bf{w}}} \right\} - \mu {{\tilde P}_m} - c.
\end{align}

The Lagrange dual function is written as
\begin{align}\label{Lagrange dual function_w_case 1}
	{l}\left( {{\mu}} \right) =& \mathop {\min }\limits_{\bf{w}} {\mathcal{L}}\left( {{\bf{w}},{\mu}} \right).
\end{align}

The corresponding Lagrangian dual problem is given by
\begin{align}\label{Lagrangian dual problem_w_case 1}
	\mathop {\max }\limits_{{\mu}}~~~& {l}\left( {{\mu}} \right)\notag\\
	{\rm{s.}}{\rm{t.}}~~~& {\mu} \ge 0.
\end{align}

Upon setting the first-order derivative of ${\mathcal{L}}\left( {{\bf{w}},{\mu}} \right)$ w.r.t. ${\bf{w}}$ to zero, the optimal solution of $\bf{w}$ is found to be 
\begin{align} \label{first-order-w case 1}
	\frac{{\partial {\mathcal{L}}\left( {{\bf{w}},{\mu}} \right)}}{{\partial {\bf{w}}^*}} = 0.
\end{align}

The left hand side of \eqref{first-order-w case 1} is recast as
\begin{align}
	2\left( {{\bf{\tilde \Xi }} + \mu {{\bf{Z}}_{\bf{\Gamma }}}} \right){\bf{w}} - 2{\bm{\omega }} - 2\mu \left( {{{\bf{Z}}_{\bf{\Gamma }}} - {\bf{\Gamma }}} \right){{\bf{w}}^{\left( t \right)}} = 0.
\end{align}

The optimal solution of $\bf{w}$ is given by
\begin{align}\label{solution of w case1}
	{\bf{w}}\left( \mu  \right) = {\left( {{\bf{\tilde \Xi }} + \mu {{\bf{Z}}_{\bf{\Gamma }}}} \right)^\dag }\left( {{\bm{\omega }} + \mu \left( {{{\bf{Z}}_{\bf{\Gamma }}} - {\bf{\Gamma }}} \right){{\bf{w}}^{\left( t \right)}}} \right),
\end{align}
where the pseudo inverse is employed since the matrix ${{\bf{\tilde \Xi }} + \mu {{\bf{Z}}_{\bf{\Gamma }}}}$ may be not full rank.

The value of $\mu$ is obtained by evaluating the complementary slack condition of constraint \eqref{replaced_Reformulation_BS2_c}, as follows
\begin{align}\label{mu}
	\mu \left( {{{\bf{w}}^H}{{\bf{Z}}_{\bf{\Gamma }}}{\bf{w}} - 2{\mathop{\rm Re}\nolimits} \left\{ {{{\left( {{{\bf{w}}^{\left( t \right)}}} \right)}^H}\left( {{{\bf{Z}}_{\bf{\Gamma }}} - {\bf{\Gamma }}} \right){\bf{w}}} \right\} - {{\tilde P}_m}} \right) = 0.
\end{align}

In order to obtain the optimal $\mu^ {\star}  \ge 0$, we need to check whether $\mu=0$ is the optimal solution or not. If the following condition holds
\begin{align}
	{\left( {{\bf{w}}\left( 0 \right)} \right)^H}{{\bf{Z}}_{\bf{\Gamma }}}{\bf{w}}\left( 0 \right) - 2{\mathop{\rm Re}\nolimits} \left\{ {{{\left( {{{\bf{w}}^{\left( t \right)}}} \right)}^H}\left( {{{\bf{Z}}_{\bf{\Gamma }}} - {\bf{\Gamma }}} \right){\bf{w}}\left( 0 \right)} \right\} \le {{\tilde P}_m},
\end{align}
for Problem \eqref{replaced_Problem_reformulaton_BS2_case1}, the optimal value of $\mu$ is $\mu^\star=0$ and the optimal transmit beamforming is ${\bf{ w}}\left(0 \right)$. Otherwise, the optimal transmit beamforming is ${\bf{ w}}\left(\mu^{\star} \right)$ with the optimal value of $\mu^{\star}$ obtained by dealing with the following equation
\begin{align}\label{solve mu}
	\!\!\!{{\tilde P}_m}\left( {{\mu ^ \star }} \right) \triangleq& {\bf{w}}{\left( {{\mu ^ \star }} \right)^H}{{\bf{Z}}_{\bf{\Gamma }}}{\bf{w}}\left( {{\mu ^ \star }} \right)\nonumber \\&- 2{\mathop{\rm Re}\nolimits} \left\{ {{{\left( {{{\bf{w}}^{\left( t \right)}}} \right)}^H}\left( {{{\bf{Z}}_{\bf{\Gamma }}} - {\bf{\Gamma }}} \right){\bf{w}}\left( {{\mu ^ \star }} \right)} \right\} = {\tilde P}_m.
\end{align}

To this end, a lemma is introduced as follows.
\begin{lemma}\label{lemma 2}
	${{\tilde P}_m}\left( {{\mu}} \right)$ is a monotonically non-increasing function of $\mu$.
\end{lemma}


Based on Lemma \ref{lemma 2}, the bisection search based approach is applied for obtaining the optimal value of $\mu$. The developed method to tackle Problem \eqref{replaced_Problem_reformulaton_BS2_case1} is illustrated in Algorithm \ref{Case1}.

\begin{algorithm}[t] 
	\caption{Bisection Search Method for Solving Probelm \eqref{replaced_Problem_reformulaton_BS2_case1}} 
	\label{Case1} 
	\textbf{Initialize}: Initialize the accuracy $\varepsilon>0$, the bounds $\mu^l$ and $\mu^u$.
	\begin{algorithmic}[1]
		\State If ${{\tilde P}_m}\left( 0 \right)\le {{\tilde P}_m}$ holds, the optimal transmit beamforming vector is obtained by ${{\bf{w}}^\star}={\bf{ w}}\left(0 \right)$ and terminate; Otherwise, go to step 2;
		\Repeat 
		\State Let $\mu = (\mu^l+\mu^u)/2$;
		\State Calculate ${\bf{w}}\left( \mu  \right)$ and ${{\tilde P}_m}\left( {{\mu}} \right)$ according to \eqref{solution of w case1} and \eqref{solve mu};
		\State If  ${{\tilde P}_m}\left( {{\mu}} \right)\ge{{\tilde P}_m}$, set $\mu^l=\mu$; Otherwise, set $\mu^u=\mu$;
		\Until $\left|\mu^u-\mu^l\right|\le\varepsilon$. Output $\mu^\star=\mu$ and  ${\bf w}^\star={\bf{w}}\left( \mu^\star  \right)$.
	\end{algorithmic}
\end{algorithm}

In each iteration of Algorithm \ref{Case1}, ${\bf{w}}(\mu)$ is calculated in \eqref{solution of w case1}, which involves a matrix inverse operation, i.e., the calculation of ${\left( {{\bf{\tilde \Xi }} + \mu {{\bf{Z}}_{\bf{\Gamma }}}} \right)^\dag }$. 
Here, we provide a new method to avoid matrix inversion operations, thereby reducing the computational complexity. Specifically, ${\bf{\tilde \Xi }}$ denotes a positive semi-definite matrix, which can be decomposed into ${\bf{\tilde \Xi }} = {{\bf{Q}}_{{\bf{\tilde \Xi }}}}{\bf{\Lambda} _{{\bf{\tilde \Xi }}}}{\bf{Q}}_{{\bf{\tilde \Xi }}}^H$ through exploiting the singular value decomposition (SVD), in which ${{\bf{Q}}_{{\bf{\tilde \Xi }}}}{\bf{Q}}_{{\bf{\tilde \Xi }}}^H = {\bf{Q}}_{{\bf{\tilde \Xi }}}^H{{\bf{Q}}_{{\bf{\tilde \Xi }}}} = {\bf{I}}_{NK}$ and ${\bf{\Lambda} _{{\bf{\tilde \Xi }}}}$ denotes a diagonal
matrix without negative entries. Then, we have ${\left( {{\bf{\tilde \Xi }} + \mu {{\bf{Z}}_{\bf{\Gamma }}}} \right)^\dag } = {{\bf{Q}}_{{\bf{\tilde \Xi }}}}\left( {{\bf{\Lambda} _{{\bf{\tilde \Xi }}}} + \mu {{\lambda} _{\bf{\Gamma }}}{\bf{I}}} \right)^\dag{\bf{Q}}_{{\bf{\tilde \Xi }}}^H$ since ${{\bf{Z}}_{\bf{\Gamma }}} = {\lambda _{\bf{\Gamma }}}{\bf{I}}$. Furthermore, the assumption ${\bf w}^\star{{\bf w}^\star}^H < P_T$ needs to be satisfied. Therefore, in each iteration, only the product of the matrices needs to be calculated, which is much less complex than computing the inverse for the same-dimension matrices.

$~\textit{2) Case \uppercase\expandafter{\romannumeral2}:}$ Assume that the BS transmit power constraint \eqref{Reformulation_BS2_b} is active at ${\bf w}^\star$, i.e., ${\bf w}^\star{{\bf w}^\star}^H = P_T$. Then, we have ${{\bf{w}}^{\star H}}{{\bf{Z}}_{\bf{\Gamma }}}{\bf{w}}^\star = {\lambda _{\bf{\Gamma }}}{P_T}$. Accordingly, the constraint \eqref{replaced_Reformulation_BS2_c} can be transformed as follows
\begin{align}\label{replaced_Reformulation_BS2_c_case2}
	2{\mathop{\rm Re}\nolimits} \left\{ {{{\left( {{{\bf{w}}^{\left( t \right)}}} \right)}^H}\left( {{{\bf{Z}}_{\bf{\Gamma }}} - {\bf{\Gamma }}} \right){\bf{w}}} \right\} \ge {{\mathord{\buildrel{\lower3pt\hbox{$\scriptscriptstyle\frown$}} 
				\over P} }_m},
\end{align}
where ${{\mathord{\buildrel{\lower3pt\hbox{$\scriptscriptstyle\frown$}} 
			\over P} }_m}\buildrel \Delta \over = {\lambda _{\bf{\Gamma }}}{P_T} - {{\tilde P}_m} $.

Correspondingly, Problem \eqref{replaced_Problem_reformulaton_BS2} is transformed as
\begin{align}\label{replaced_Reformulation_BS2_case2}
	\mathop {{\rm{min}}}\limits_{{{\bf{w}}}}~~~&{{{\bf{w}}^H}{\bf{\tilde \Xi w}}  -2{\rm{Re}}\left\{ {{{\bm{\omega }}^H}{\bf{w}}} \right\} - c}\notag\\
	{\rm{s.}}{\rm{t.}}~~~&
	\eqref{Reformulation_BS2_b},~
	\eqref{replaced_Reformulation_BS2_c_case2}.
\end{align}

Firstly, upon introducing the Lagrange multiplier $\lambda_1$ associated with the transmit power constraint of the BS, the partial Lagrange function of Problem \eqref{Problem_reformulaton_BS2} is transformed as
\begin{align}
	{\mathcal{L}}_1\left( {{\bf{w}},{\lambda _1}} \right) =& {{\bf{w}}^H}\left( {{\bf{\tilde \Xi }} + {\lambda _1}{\bf{I}}} \right){\bf{w}} - 2{\rm{Re}}\left\{ {{{\bm{\omega }}^H}{\bf{w}}} \right\} \nonumber\\&- {\lambda _1}{P_T} - c.
\end{align}

The Lagrange dual function is expressed as
\begin{align}\label{Lagrange dual function_w}
		{l_1}\left( {{\lambda _1}} \right) = \mathop {\min }\limits_{\bf{w}}~~& {\mathcal{L}}_1\left( {{\bf{w}},{\lambda _1}} \right)~~~\notag\\
		{\rm{s.}}{\rm{t.}}~~& \eqref{replaced_Reformulation_BS2_c_case2}.
\end{align}

The corresponding Lagrangian dual problem can be written as
\begin{align}\label{Lagrangian dual problem_w}
	\mathop {\max }\limits_{{\lambda _1}}~~~ &{l_1}\left( {{\lambda _1}} \right)\notag\\
	{\rm{s.}}{\rm{t.}}~~~& {\lambda _1} \ge 0.
\end{align}

To deal with the dual Problem \eqref{Lagrangian dual problem_w}, the expression of dual function $l_1(\lambda_1)$ needs to be derived by solving Problem \eqref{Lagrange dual function_w} with given $\lambda_1$. By keeping the value of $\lambda_1$ fixed, then introducing another Lagrange multiplier $\lambda_2 \ge 0$ associated with the constraint \eqref{replaced_Reformulation_BS2_c_case2}, the Lagrange function of Problem \eqref{Lagrange dual function_w} is expressed as
\begin{align} \label{Lagrange function_lambda2}
	{\mathcal{L}}_2\left( {{\bf{w}},{\lambda _2}} \right) =& {{\mathcal{L}}_1}\left( {{\bf{w}},{\lambda _1}} \right) + {\lambda _2}{{\mathord{\buildrel{\lower3pt\hbox{$\scriptscriptstyle\frown$}} \over P} }_m}\nonumber\\&- 2{\lambda _2}{\mathop{\rm Re}\nolimits} \left\{ {{{\left( {{{\bf{w}}^{\left( t \right)}}} \right)}^H}\left( {{{\bf{Z}}_{\bf{\Gamma }}} - {\bf{\Gamma }}} \right){\bf{w}}} \right\}.
\end{align}

By setting the first-order derivative of ${\mathcal{L}}_2\left( {{\bf{w}},{\lambda _2}} \right)$ w.r.t. ${\bf{w}}$ to zero, the optimal solution of $\bf{w}$ is found as 
\begin{align} \label{first-order-w}
	\frac{{\partial {\mathcal{L}_2}\left( {{\bf{w}},{\lambda _2}} \right)}}{{\partial {\bf{w}}}} = 0.
\end{align}

The left hand side of \eqref{first-order-w} is recast as
\begin{align}
	\frac{{\partial {\mathcal{L}_2}\left( {{\bf{w}},{\lambda _2}} \right)}}{{\partial {\bf{w}}}}=2\left( {{\bf{\tilde \Xi }} + {\lambda _1}{\bf{I}}} \right){\bf{w}} - 2{\bm{\omega }} - 2{\lambda _2}\left( {{{\bf{Z}}_{\bf{\Gamma }}} - {\bf{\Gamma }}} \right){{\bf{w}}^{\left( t \right)}}.
\end{align}

Equation \eqref{first-order-w} becomes
\begin{align}
	\left( {{\bf{\tilde \Xi }} + {\lambda _1}{\bf{I}}} \right){\bf{w}} = {\bm{\omega }} + {\lambda _2}\left( {{{\bf{Z}}_{\bf{\Gamma }}} - {\bf{\Gamma }}} \right){{\bf{w}}^{\left( t \right)}}.
\end{align}

Then the solution of ${\bf{w}}(\lambda_1,\lambda_2)$ is obtained by
\begin{align}\label{solution of w}
	{\bf{w}}\left( {{\lambda_1,\lambda _2}} \right) = {\left( {{\bf{\tilde \Xi }} + {\lambda _1}{\bf{I}}} \right)^\dag }\left( {{\bm{\omega }} + {\lambda _2}\left( {{{\bf{Z}}_{\bf{\Gamma }}} - {\bf{\Gamma }}} \right){{\bf{w}}^{\left( t \right)}}} \right),
\end{align}
where we use the pseudo inverse since $ {{\bf{\tilde \Xi }} + {\lambda _1}{\bf{I}}  }$ may be not full rank.

For a given $\lambda_1$, the optimal value of $\lambda_2$ can be obtained to ensure that the complementary slack condition for constraint \eqref{replaced_Reformulation_BS2_c_case2} is satisfied:
\begin{align}\label{lambda_2}
	\!\!\!{\lambda _2^\star}\left( {2{\mathop{\rm Re}\nolimits} \left\{ {{{\left( {{{\bf{w}}^{\left( t \right)}}} \right)}^H}\left( {{{\bf{Z}}_{\bf{\Gamma }}} \!-\! {\bf{\Gamma }}} \right){\bf{w}}\left( {{\lambda _1},{\lambda _2^\star}} \right)} \right\} \!-\! {{\mathord{\buildrel{\lower3pt\hbox{$\scriptscriptstyle\frown$}} 
					\over P} }_m}} \right) = 0.
\end{align}

In order to obtain the optimal $\lambda _2^ {\star}  \ge 0$, we need to examine whether $\lambda_2=0$ is the optimal solution or not. If the following equation holds, 
\begin{align}\label{lambda2=0}
	2{\mathop{\rm Re}\nolimits} \left\{ {{{\left( {{{\bf{w}}^{\left( t \right)}}} \right)}^H}\left( {{{\bf{Z}}_{\bf{\Gamma }}} - {\bf{\Gamma }}} \right){\bf{w}}\left( {{\lambda _1},0} \right)} \right\} \ge {{\mathord{\buildrel{\lower3pt\hbox{$\scriptscriptstyle\frown$}} 
				\over P} }_m},
\end{align}
the optimal value of $\lambda_2$ is $\lambda_2^\star=0$ and the optimal transmit beamforming is ${\bf{ w}}\left(\lambda_1, 0 \right)$. Otherwise, the optimal transmit beamforming is ${\bf{ w}}\left(\lambda_1, \lambda_2^{\star} \right)$ with the optimal value $\lambda_2^{\star}$, which is derived as
\begin{align}\label{solve lambda_2}
	\lambda_2^\star= \frac{{{{\mathord{\buildrel{\lower3pt\hbox{$\scriptscriptstyle\frown$}} \over P} }_m} - 2{{\left( {{{\bf{w}}^{\left( t \right)}}} \right)}^H}\left( {{{\bf{Z}}_{\bf{\Gamma }}} - {\bf{\Gamma }}} \right){{\left( {{\bf{\tilde \Xi }} + {\lambda _1}{\bf{I}}} \right)}^\dag }{\bm{\omega }}}}{{2{{\left( {{{\bf{w}}^{\left( t \right)}}} \right)}^H}\left( {{{\bf{Z}}_{\bf{\Gamma }}} - {\bf{\Gamma }}} \right){{\left( {{\bf{\tilde \Xi }} + {\lambda _1}{\bf{I}}} \right)}^\dag }\left( {{{\bf{Z}}_{\bf{\Gamma }}} - {\bf{\Gamma }}} \right){{\bf{w}}^{\left( t \right)}}}}.
\end{align}

With the expression of the dual function $l_1(\lambda_1)$, we consider Problem \eqref{Lagrangian dual problem_w} to find the optimal $\lambda_1$. The optimal value of $\lambda_1$ is the solution to the dual Problem \eqref{Lagrangian dual problem_w}, which also has to meet the complementary slackness condition of the BS transmit power constraint. Moreover, the optimal value of $\lambda_2$ has been derived through the above calculations. Then, $\lambda_1^{\star}$ is found by solving
\begin{align}
	\lambda _1^\star\left( {{{\left( {{\bf{w}}\left( {\lambda _1^\star,\lambda _2^\star} \right)} \right)}^H}{\bf{w}}\left( {\lambda _1^\star,\lambda _2^\star} \right) - {P_T}} \right) = 0.
\end{align}

Similarly, to obtain the optimal $\lambda _1^ {\star}  \ge 0$, we have to check whether $\lambda_1=0$ is the optimal solution or not. If the following inequality is met, 
\begin{align}
	{\left( {{\bf{w}}\left( {0,\lambda _2^\star} \right)} \right)^H}{\bf{w}}\left( {0,\lambda _2^\star} \right) \le {P_T},
\end{align}
the optimal value of $\lambda_1$ is $\lambda_1^\star=0$. Otherwise, the optimal value of $\lambda_1$ is found through solving the following equation:
\begin{align}\label{solve lambda_1}
	{P_T}\left( {{\lambda _1^\star}} \right){\rm{ }} \buildrel \Delta \over = {\left( {{\bf{w}}\left( {\lambda _1^ \star ,\lambda _2^ \star } \right)} \right)^H}{\bf{w}}\left( {\lambda _1^ \star ,\lambda _2^ \star } \right) = P_T.
\end{align}

\begin{algorithm}[t] 
	\caption{Bisection Search Method for Solving Probelm \eqref{replaced_Reformulation_BS2_case2}} 
	\label{Case2} 
	
	\textbf{Initialize}: Initialize the accuracy $\varepsilon>0$, the bounds $\lambda_1^l$ and $\lambda_1^u$.
	\begin{algorithmic}[1]
		\Repeat 
		\State Obtain $\lambda_1 = (\lambda_1^l+\lambda_1^u)/2$;
		\State If \eqref{lambda2=0} holds, $\lambda_2^\star=0$; Otherwise, $\lambda_2^\star$ is given in \eqref{solve lambda_2};
		\State Update ${\bf{w}}\left( {{\lambda_1,\lambda _2^\star}} \right)$ according to \eqref{solution of w};
		\State If ${\left( {{\bf{w}}\left( {0,\lambda _2^\star} \right)} \right)^H}{\bf{w}}\left( {0,\lambda _2^\star} \right) \le {P_T}$ holds, $\lambda_1^\star=0$; Otherwise, go to step 6;
		\State If ${{ P}_T}\left( {{\lambda_1}} \right)\ge{{ P}_T}$, set $\lambda_1^l=\lambda_1$; Otherwise, set $\lambda_1^u=\lambda_1$;
		\Until $\left|\lambda_1^u-\lambda_1^l\right|\le\varepsilon$. Output $\lambda_1^\star=\lambda_1$, $\lambda_2^\star=\lambda_2$ and ${\bf w}^\star={\bf{w}}\left( \lambda_1^\star,\lambda_2^\star \right)$.
	\end{algorithmic}
\end{algorithm}

To tackle Problem \eqref{solve lambda_1}, Lemma \ref{lemma 3} is provided as follows.
\begin{lemma}\label{lemma 3}
	${{ P}_T}\left( {{\lambda_1}} \right)$ is a monotonically non-increasing function of $\lambda_1$.
\end{lemma}


We proved that ${P_T}\left( {{\lambda _1}} \right)$ is monotonically non-increasing w.r.t. $\lambda_1$ in Lemma \ref{lemma 3}, which enables the bisection search approach to be employed for obtaining the solution to Problem \eqref{solve lambda_1}. Moreover, the description of the developed approach for solving Problem \eqref{replaced_Reformulation_BS2_case2} is detailed in Algorithm \ref{Case2}.

Based on the above
discussion, the overall algorithm to tackle Problem \eqref{Problem_reformulaton_BS2} is summarized in Algorithm \ref{Bisection Search}.

In the following lemma, we provide the convergence of Algorithm \ref{Bisection Search}.

\begin{lemma}\label{lemma 4}
	The sequence $\{{\bf{w}}^{(n)}, n=1,2,\cdots\}$ obtained by Algorithm \ref{Bisection Search} converges to the KKT optimum point
	of Problem \eqref{Problem_reformulaton_BS2}.
\end{lemma}
%

\subsection{Optimize the RIS Reflection Coefficient Matrix ${\bf{\Psi }}$}
In this subsection, we start to design the active RIS reflection coefficient matrix while the other variables are fixed. The subproblem in \eqref{Problem_reformulaton_RIS2} is an SOCP problem, which can be addressed by exploiting CVX tools.  To reduce the computational complexity, we adopt the low-complexity method based on MM framework. To this end, a lemma is introduced as below.

\begin{algorithm}[t] 
	\caption{MM-based Lagrangian dual decomposition Algorithm for Solving Probelm \eqref{Problem_reformulaton_BS2}} 
	\label{Bisection Search} 
	
	\textbf{Initialize}: Initialize feasible ${\bf{w}}^{(0)}$ and given ${\bm{\psi}}$, set the iteration number $n = 0$.
	\begin{algorithmic}[1]
		\Repeat 
		\State Obtain ${\bf{w}}^{(n)}$ by solving Problem \eqref{replaced_Problem_reformulaton_BS2_case1} by using Algorithm \ref{Case1};
		\State If ${\bf{w}}^{(n)}$ satisfies constraint \eqref{Reformulation_BS2_b}, go to step 5;
		\State Obtain ${\bf{w}}^{(n)}$ by solving Problem \eqref{replaced_Reformulation_BS2_case2} by using Algorithm \ref{Case2};
		\State Set $n \gets n+1$;
		\Until The value of the objective function in \eqref{Problem_reformulaton_BS2} converges.
	\end{algorithmic}
\end{algorithm}

\begin{lemma}\label{lemma 5}
	Let ${f_{\bf{\Delta }}}\left( {\bf{\Delta }} \right) = {{\bm{\psi}}^H}{\bf{\Delta}} {\bm \psi}$, ${\bf{\Delta }} \succeq \bf{0}$, ${{\bf{Z}}_{\bf{\Delta }}} = {\lambda _{\bf{\Delta }}}{\bf{I}}_M$, where ${\lambda _{\bf{\Delta }}}$ represents the maximum eigenvalue of ${\bf{\Delta }}$. Then, for any given solution ${{\bm{\psi}}^{\left( t \right)}}$ at the $t$-th iteration and for any feasible ${\bm{\psi}}$, we have
	\begin{align}\label{lemma8}
		{{\tilde f}_{\bf{\Delta }}}\left( {{\bm{\psi}}\left| {{{\bm{\psi}}^{\left( t \right)}}} \right.} \right) =& {{\bm{\psi}}^H}{{\bf{Z}}_{\bf{\Delta }}}{\bm{\psi}} + {\left( {{{\bm{\psi}}^{\left( t \right)}}} \right)^H}\left( {{{\bf{Z}}_{\bf{\Delta }}} - {\bf{\Delta }}} \right){{\bm{\psi}}^{\left( t \right)}} \nonumber\\&- 2{\mathop{\rm Re}\nolimits} \left\{ {{{\left( {{{\bm{\psi}}^{\left( t \right)}}} \right)}^H}\left( {{{\bf{Z}}_{\bf{\Delta }}} - {\bf{\Delta }}} \right){\bm{\psi}}} \right\},
	\end{align}
\begin{equation}
	~\notag
\end{equation}
	which satisfies three conditions as follows:
	\begin{enumerate}
		\item [1)] 
		${{\tilde f}_{\bf{\Delta }}}\left( { {{{\bm{\psi}}^{\left( t \right)}}} \left| {{{\bm{\psi}}^{\left( t \right)}}} \right.} \right) = {f_{\bf{\Delta }}}\left( {{{\bm{\psi}}^{\left( t \right)}}} \right)$;      
		\item [2)]
		${\nabla _{\bm{\psi}}}{{\tilde f}_{\bf{\Delta }}}{\left( {{\bm{\psi}}\left| {{{\bm{\psi}}^{\left( t \right)}}} \right.} \right)_{{\bm{\psi}} = {{\bm{\psi}}^{\left( t \right)}}}} = {\nabla _{\bm{\psi}}}{f_{\bf{\Delta }}}{\left( {{{\bm{\psi}}^{\left( t \right)}}} \right)_{{\bm{\psi}} = {{\bm{\psi}}^{\left( t \right)}}}}$;
		\item [3)]
		${{\tilde f}_{\bf{\Delta}}}\left( {{\bm{\psi}}\left| {{{\bm{\psi}}^{\left( t \right)}}} \right.} \right) \ge {f_{\bf{\Delta }}}\left( {\bm{\psi}} \right)$.
	\end{enumerate}
\end{lemma}

Similar to \eqref{lemma8}, we have 
\begin{align}
	{{\tilde f}_{\bf{\Lambda }}}\left( {{\bm{\psi}}\left| {{{\bm{\psi}}^{\left( t \right)}}} \right.} \right) =& {{\bm{\psi}}^H}{{\bf{Z}}_{\bf{\Lambda }}}{\bm{\psi}} + {\left( {{{\bm{\psi}}^{\left( t \right)}}} \right)^H}\left( {{{\bf{Z}}_{\bf{\Lambda }}} - {\bf{\Lambda }}} \right){{\bm{\psi}}^{\left( t \right)}} \nonumber\\&- 2{\mathop{\rm Re}\nolimits} \left\{ {{{\left( {{{\bm{\psi}}^{\left( t \right)}}} \right)}^H}\left( {{{\bf{Z}}_{\bf{\Lambda }}} - {\bf{\Lambda }}} \right){\bm{\psi}}} \right\}.\label{method2_psi}
\end{align}

For a given ${{{\bm{\psi}}^{\left( t \right)}}}$, combining \eqref{lemma8} and \eqref{method2_psi}, Problem \eqref{Problem_reformulaton_RIS2} is rewritten as
\begin{subequations}\label{Problem_reformulaton_RIS3}
	\begin{align}
		\mathop {{\rm{min}}}\limits_{{{\bm{\psi}}}}~~~&f\left( {\bm \psi}  \right) \buildrel \Delta \over = {{\bm{\psi}} ^H}{{\bf{Z}}_{\bf{\Delta }}}{\bm{\psi}}  + {\mathop{\rm Re}\nolimits} \left\{ {{{\bm{\psi }}^H}{\bf{\tilde p}}} \right\} + \tilde d \label{Reformulation_RIS_a3}\\ \label{Reformulation_RIS_b3}
		{\rm{s.}}{\rm{t.}}~~~&g\left( {\bm \psi}  \right) \buildrel \Delta \over = {{\bm\psi} ^H}{{\bf{Z}}_{\bf{\Lambda }}}{\bm\psi}  + {\rm{Re}}\left\{ {{{\bm\psi} ^H}{\bf{\tilde q}}} \right\} \le {{\tilde P}_A},
	\end{align}
\end{subequations}
where ${\bf{\tilde p}} \triangleq  - 2\left( {{\bm{\alpha }} + {{\left( {{{\bf{Z}}_{\bf{\Delta }}} - {\bf{\Delta }}} \right)}^H} {{{\bm{\psi}} ^{\left( t \right)}}} } \right)$, $\tilde d \triangleq -d + {\left( {{{\bm{\psi}} ^{\left( t \right)}}} \right)^H}\left( {{{\bf{Z}}_{\bf{\Delta }}} - {\bf{\Delta }}} \right){{\bm{\psi}} ^{\left( t \right)}}$,  ${\bf{\tilde q}} \triangleq -2{\left( {{{\bf{Z}}_{\bf{\Lambda }}} - {\bf{\Lambda }}} \right)^H}{{\bm\psi} ^{\left( t \right)}}$ and ${{\tilde P}_A} \triangleq {P_A} - {\left( {{{\bm{\psi}} ^{\left( t \right)}}} \right)^H}\left( {{{\bf{Z}}_{\bf{\Lambda }}} - {\bf{\Lambda }}} \right){{\bm{\psi}} ^{\left( t \right)}}$.

Based on these considerations, we develop a price mechanism, through which Problem \eqref{Problem_reformulaton_RIS3} can be equivalently reformulated as the following form
\begin{equation}\label{Price mechanism}
		\mathop {{\rm{min}}}\limits_{{{\bm{\psi}}}}~~~h\left( {\bm \psi}  \right) \buildrel \Delta \over = f\left( {\bm \psi}  \right) + \eta g\left( {\bm \psi}  \right) 
\end{equation}
where $\eta \ge 0$ stands for the introduced price of the function $g\left( {\bm \psi}  \right)$.


The value of $\eta$ can be found by exploiting the subgradient approach or the bisection search method similar to Algorithm \ref{Case1}. Thereby, the detailed procedures are omitted here.

Subsequently, for a given $\eta$, we conceive an effective element-wise ASO approach to solve Problem \eqref{Price mechanism} for obtaining a high-quality suboptimal solution of the RIS reflection coefficient vector ${\bm{\psi }}$. 

\begin{algorithm}[t] 
	\caption{MM-based Element-wise ASO Algorithm for Solving Probelm \eqref{Price mechanism}} 
	\label{ASO} 
	
	\textbf{Initialize}: Initialize feasible ${\bm{a}}^{(0)}$, ${\bm{\theta}}{^{(0)}}$ and given ${\bf{w}}$, set the iteration number $n = 0$.
	\begin{algorithmic}[1]
		\Repeat 
		\State Obtain the price $\eta$ by bisection
		search approach;
		\State Alternately optimize $\theta_m^{(n)}$ and $a_m^{(n)}$ according to \eqref{solution of theta} and \eqref{solution of am};
		\State Obtain ${\bm{a}}^{(n)} \buildrel \Delta \over = {\left[ {{a_1^{(n)}}, \cdots ,{a_M^{(n)}}} \right]^T}$, ${\bm{\theta }}^{(n)} \buildrel \Delta \over = {\left[ {{{{\theta _1^{(n)}}}}, \cdots ,{{{\theta _M^{(n)}}}}} \right]^T}$ and ${\bm \psi}^{(n)}  \buildrel \Delta \over = {\bm{a}}^{(n)} \odot {\bm{\theta }}^{(n)}$;
		\State Set $n \gets n+1$;
		\Until The value of the objective function $h\left( {\bm \psi}  \right)$ in \eqref{Price mechanism} converges.
	\end{algorithmic}
\end{algorithm}

Recalling that ${\bf{\Psi }} = {\bf{A\Theta }} = {\rm{diag}}\left[ {{a_1}{e^{j{\varphi _1}}}, \cdots ,{a_M}{e^{j{\varphi _M}}}} \right] \in {{\mathbb{C}}^{M \times M}}$, we define ${\bm \psi}  \buildrel \Delta \over = {\bm{a}} \odot {\bm{\theta }}$ as the diagonal elements of matrix ${\bf{\Psi}}$, where ${\bm{a}} \buildrel \Delta \over = {\left[ {{a_1}, \cdots ,{a_M}} \right]^T}$ and ${\bm{\theta }} \buildrel \Delta \over = {\left[ {{{{\theta _1}}}, \cdots ,{{{\theta _M}}}} \right]^T}$. More specifically, the objective function in \eqref{Price mechanism} is further recast as an equivalent function w.r.t. $a_m$ and $\theta_m$, which is rewritten as
\begin{align}\label{62}
	h\left( {{a_m},{\theta _m}} \right)
	=& \left( {{{\tilde z}_{m,m}} \!+\! \eta {{\mathord{\buildrel{\lower3pt\hbox{$\scriptscriptstyle\frown$}} \over z} }_{m,m}}} \right)a_m^2 + {\mathop{\rm Re}\nolimits} \left\{ {\left( {{{{{\tilde p}}}_m} \!+\! \eta {{{{\tilde q}}}_m}} \right)\theta _m^*{a_m}} \right\} \nonumber\\&+ \sum\limits_{i = 1,i \ne m}^M {\left( {{{\tilde z}_{i,i}} + \eta {{\mathord{\buildrel{\lower3pt\hbox{$\scriptscriptstyle\frown$}} \over z} }_{i,i}}} \right)a_i^2} \nonumber\\& + {\mathop{\rm Re}\nolimits} \left\{ {\sum\limits_{i = 1,i \ne m}^M {\left( {{{{{\tilde p}}}_i} + \eta {{{{\tilde q}}}_i}} \right){a_i}} } \right\} + \tilde d,
\end{align}
where ${{\tilde z}_{m,m}}$ and ${{\mathord{\buildrel{\lower3pt\hbox{$\scriptscriptstyle\frown$}} \over z} }_{m,m}}$ are the $m$-th diagonal entry of the diagonal matrices ${{\bf{Z}}_{\bf{\Delta }}}$ and ${{\bf{Z}}_{\bf{\Lambda }}}$, respectively. As a result, we only need to explore the following problem to sequentially optimize a pair of variables $\{a_m,\theta_m\}$ while keeping the other $\left( M-1\right)$ pairs fixed. The constant terms in $h\left( {{a_m},{\theta _m}} \right)$ are omitted, which makes no difference in optimizing $a_m$ and $\theta_m$. Defining ${b_m} = {{\tilde p}_m} + \eta {{\tilde q}_m}$, Problem \eqref{Price mechanism} can be recast as
\begin{subequations}\label{an in Price mechanism}
	\begin{align}
	\mathop {{\rm{min}}}\limits_{{{a_m,\theta_m}}}~~~&\left( {{{\tilde z}_{m,m}} + \eta {{\mathord{\buildrel{\lower3pt\hbox{$\scriptscriptstyle\frown$}} \over z} }_{m,m}}} \right)a_m^2 + {\mathop{\rm Re}\nolimits} \left\{ {{b_m}\theta _m^*{a_m}} \right\} \label{an OF in price}\\ {\rm{s.}}{\rm{t.}}~~~~&\angle {\theta _m}=\varphi_m \in \left[ {0,2\pi } \right] \label{angel constraint},
	\end{align}
\end{subequations}
where the constraint in  \eqref{angel constraint} is the angle constraint at each phase shift.

Noting that only $\theta_m$ appears in the constraint of \eqref{angel constraint}, Problem \eqref{an in Price mechanism} can be decomposed into two subproblems as below
\begin{subequations}\label{12an in Price mechanism}
	\begin{align}
		\mathop {{\rm{min}}}\limits_{{{\theta_m}}}~~~&{\mathop{\rm Re}\nolimits} \left\{ {{b_m}\theta _m^*{a_m}} \right\} \label{2an OF in price}\notag\\
		&{\rm{s.}}{\rm{t.}}~\eqref{angel constraint},\\ \mathop {{\rm{min}}}\limits_{{a_m}}~~~&\left( {{{\tilde z}_{m,m}} + \eta {{\mathord{\buildrel{\lower3pt\hbox{$\scriptscriptstyle\frown$}} \over z} }_{m,m}}} \right)a_m^2 + {\mathop{\rm Re}\nolimits} \left\{ {{b_m}\theta _m^*{a_m}} \right\} \label{1an OF in price}
	\end{align}
\end{subequations}

\begin{algorithm}[t] 
	\caption{ASO-based BCD Algorithm for Solving Probelm \eqref{Problem_reformulaton}} 
	\label{BCD-ASO} 
	
	\textbf{Initialize}: Initialize feasible ${\bf{w}}^{(0)}$, ${\bm{\psi}}^{(0)}$, error tolerance $\varepsilon$, maximum number of iterations $n_\max$, set the iteration number $n = 0$, evaluate the objective function value of Problem \eqref{Problem_reformulaton}, denoted as ${\text{Obj}}({\bf{w}}^{(0)},{\bm{\psi}}^{(0)})$.
	\begin{algorithmic}[1]
		\Repeat 
		\State Given ${\bf{w}}^{(n)}$ and ${\bm{\psi}}^{(n)}$, evaluate the auxiliary variables ${\mathcal{U}}^{(n+1)}$ in \eqref{optimal_u} and $\mathcal{V}^{(n+1)}$ in \eqref{optimal_v};
		\State Given ${\bm{\psi}}^{(n)}$, $\mathcal{U}^{(n+1)}$ and ${\mathcal{V}}^{(n+1)}$, obtain ${\bf{w}}^{(n+1)}$ by solving Problem \eqref{Problem_reformulaton_BS2} by using the Lagrangian multiplier approach given in Algorithm \ref{Bisection Search};
		\State Given ${\bf{w}}^{(n+1)}$, $\mathcal{U}^{(n+1)}$ and ${\mathcal{V}}^{(n+1)}$, obtain ${\bm{\psi}}^{(n+1)}$ by solving Problem \eqref{Price mechanism} with element-wise ASO method detailed in Algorithm \ref{ASO};
		\State Set $n \gets n+1$;
		\Until $n > n_\max$ or  $\frac{{\text{Obj}}({\bf{w}}^{(n+1)},{\bm{\psi}}^{(n+1)})-{\text{Obj}}({\bf{w}}^{(n)},{\bm{\psi}}^{(n)})}{{\text{Obj}}({\bf{w}}^{(n)},{\bm{\psi}}^{(n)})}<\varepsilon .$
	\end{algorithmic}
\end{algorithm}

\begin{figure}[t]
	\begin{center}
		\includegraphics[scale=0.7]{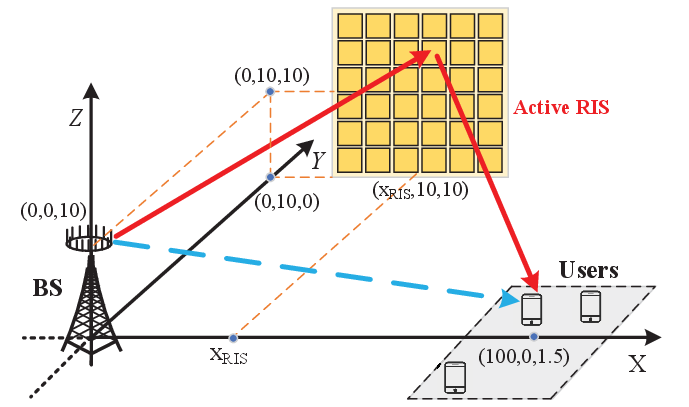}
		\caption{Simulation setup of active RIS-assisted multiuser MISO communication system.}
		\label{simulation}
	\end{center}
\end{figure}

In particular, Problem \eqref{2an OF in price} can be equivalently transformed into
\begin{subequations}
	\begin{align}
	\mathop {\min }\limits_{{\varphi _m}}~~~& \cos \left( { - {\varphi _m} + \angle {b_m}} \right)\label{cos}\\
	{\rm{s}}{\rm{.t}}{\rm{.}}~~~&{\varphi _m} \in \left[ {0,2\pi } \right].
	\end{align}
\end{subequations}

For a given $b_m$, the globally optimal solution can be obtained as
\begin{align}
	{\varphi _m} = \angle {b_m} - \pi .
\end{align}

Correspondingly, $\theta_m$ is obtained as
\begin{align}\label{solution of theta}
	{\theta _m} = {e^{j\left( {\angle {b_m} - \pi } \right)}}.
\end{align}

Next, we consider Problem \eqref{1an OF in price}. With the obtained
solution of $\theta_m$, the value of \eqref{cos}
is calculated as $-1$ and consequently, ${\mathop{\rm Re}\nolimits} \left\{ {{b_m}\theta _m^*{a_m}} \right\}$ can be calculated as $ - \left| {{b_m}} \right|{a_m}$. Therefore, the solution of the unconstrained Problem \eqref{1an OF in price} can be found by the following closed-form expression
\begin{align}
	{a_m} = \frac{\left| {{b_m}} \right|}{{2\left( {{{\tilde z}_{m,m}} + \eta {{\mathord{\buildrel{\lower3pt\hbox{$\scriptscriptstyle\frown$}} \over z} }_{m,m}}} \right)}}.\label{solution of am}
\end{align}


The detailed procedure of the proposed element-wise ASO algorithm for solving Problem \eqref{Price mechanism} is shown in Algorithm \ref{ASO}. It can be proved that Algorithm \ref{ASO} converges as elaborated in \cite{7946256}.

Based on the above discussion, the proposed BCD-ASO algorithm to solve Problem \eqref{Problem_reformulaton} is presented in Algorithm \ref{BCD-ASO}.

\begin{figure}[t]
	\begin{center}
		\includegraphics[scale=0.65]{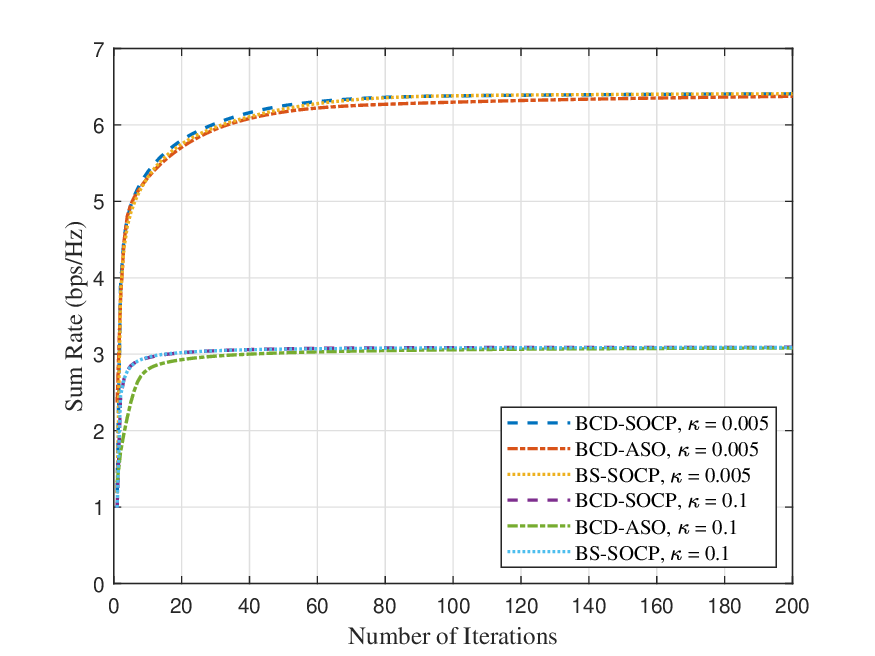}
		\caption{Convergence behaviour of the proposed algorithms. 
		}
		\label{iteration}
	\end{center}
\end{figure}

\begin{figure}[t]
	\begin{center}
		\includegraphics[scale=0.65]{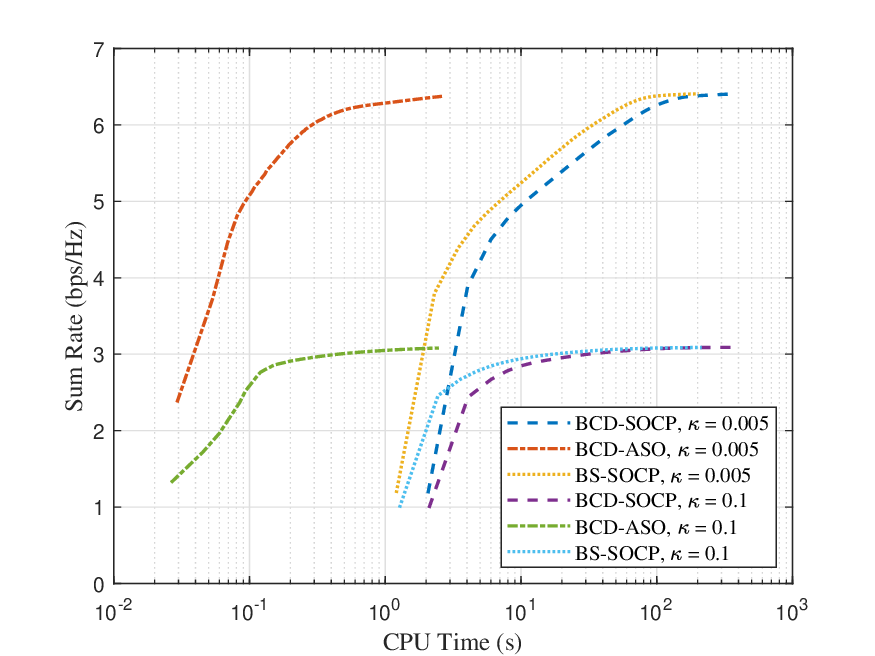}
		\caption{CPU time of the proposed algorithms. 
		}
		\label{CPUTime}
	\end{center}
\end{figure}

\section{Simulation Results}\label{Section 5}
In this section, representative simulation results are provided to validate the benefits of employing an active RIS for improving the performance of the considered multiuser MISO communication system in the presence of HWIs. The performance of the designed algorithms is evaluated as well.

\subsection{Simulation Setup}

Consider a three-dimensional (3D) coordinate setup as depicted in Fig. \ref{simulation}, in which 
the BS and the active RIS are situated at $(0,0,10)$ and $(80,10,10)$ in meters, respectively. All users are randomly distributed within a disk centered at $(100,0,1.5)$. All the channels follow a Rician distribution. Unless otherwise specified, the simulation parameters are set as follows: the number of RIS elements $M = 16$, the number of BS transmit antennas $N = 4$, the number of users $K = 3$, the T-HWI coefficients $\kappa_t = \kappa_{r}= \kappa = 0.01^2$, the amplifier power coefficients ${\xi _T} = {\xi _A} = 1.2$, the circuit dissipated power at BS $P_{\rm{BS}} = 9$ dBW \cite{8741198}, the power budget $P = 20$ dBm. \footnote{The hardware static power at BS $P_{\rm{BS}}$ exists in all systems, it thus has no impact on the comparison results between active RIS and passive RIS. Therefore, in simulation, we denote the power budget as $P$, which does not include $P_{\rm{BS}}$.}


%

\subsection{Baseline Schemes}
Algorithm \ref{SOCP Algorithm} in Section \ref{Section 3} is denoted as $\textbf{BCD-SOCP}$ and Algorithm \ref{BCD-ASO} in Section \ref{Section 4} is denoted as $\textbf{BCD-ASO}$. In this section, the simulations are performed for comparing the performance of the developed algorithms with six baseline schemes as follows:


\begin{itemize}
	\item [1)] $\textbf{BS-SOCP:~}$We design a benchmark algorithm, where the BS beamforming matrix is designed by bisection search method in Algorithm \ref{Bisection Search} 
	and the reflection coefficients of the active RIS are obtained by Algorithm \ref{SOCP Algorithm}. 
	
	\item [2)] $\textbf{Active RIS \cite{zhang2021active}:~}$ In the same scenario as considered in this paper, the algorithm proposed in \cite{zhang2021active} is adopted to jointly optimize the beamforming of the BS and the reflection coefficients of the active RIS.
	
	\item [3)] $\textbf{Active RandPhase:~}$
	The phase shifts of the active RIS are randomly generated from $[0,2\pi]$. The BS beamforming matrix and the amplification factors are optimized by Algorithm \ref{Bisection Search} and Algorithm \ref{ASO}, respectively.
	
	\item [4)] $\textbf{Passive RIS \cite{9090356}:~}$The traditional nearly passive RIS without amplification is considered as another benchmark. The algorithm proposed in \cite{9090356} is used to jointly optimize the beamforming of the BS and the phase shifts of the passive RIS. In addition, for a fair comparison, the total power budget of the nearly passive RIS-assisted system is assumed to be the same as that of active RIS system.\footnote{In nearly passive RIS system, there is no DC biasing power consumption at each reconfigurable element, and we have $P = {\xi _T}P_T + M P_{\rm{SW}}$.}
	
	\item [5)] $\textbf{Passive RandPhase \cite{5756489}:~}$ The phase shifts of the passive RIS are randomly generated, similar to scheme 3). The weighted minimum mean square error (WMMSE) method proposed in \cite{5756489} is adopted to optimize the beamforming of the BS.
	
	\item [6)] $\textbf{Without RIS \cite{5756489}:~}$There is no RIS to aid the transmission. The WMMSE method from \cite{5756489} is used to design the beamforming of the BS. Similar to scheme 4), the total power budget keeps the same as that of the active RIS-assisted system.\footnote{In the system without RIS, the power budget is used as transmit power of the BS, and we have $P = {\xi _T}P_T$.}
\end{itemize}

\begin{figure}[t]
	\vspace{0.2cm}	
	\begin{center}
		\includegraphics[scale=0.65]{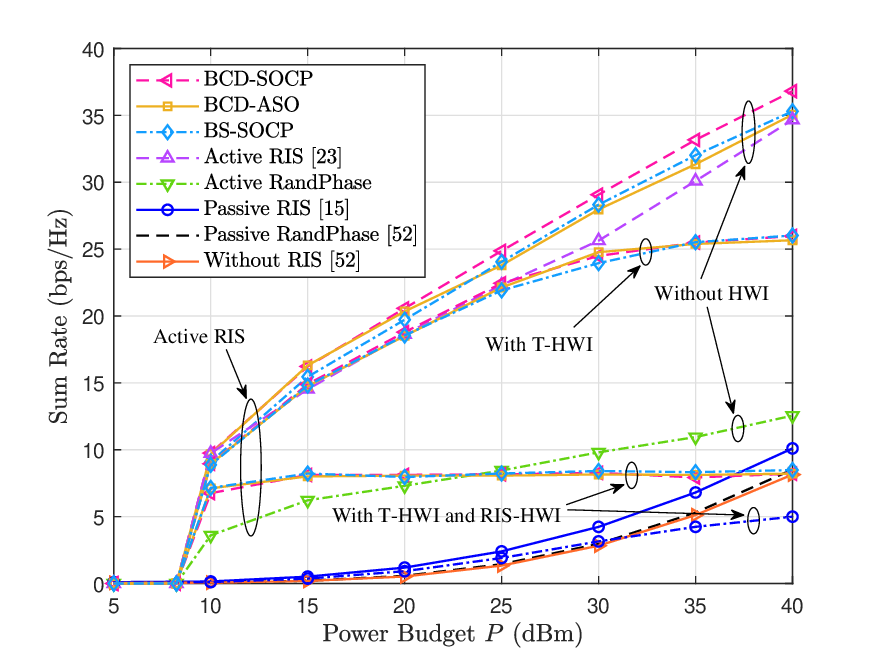}
		\caption{Sum rate versus the power budget $P$.
		}
		\label{power}
	\end{center}
\end{figure}

\subsection{Convergence of the Proposed Algorithms}

Fig. \ref{iteration} and Fig. \ref{CPUTime} show the convergence behavior and the CPU time of various schemes while the power budget is $P= 10$ dBm. Fig. \ref{iteration} shows that all the schemes can reach convergence within 100 iterations, which demonstrates the effectiveness of the presented algorithms. The BCD-ASO algorithm achieves almost the same performance as the other two SOCP-based algorithms in terms of sum rate, while having a significant advantage in computational time owing to the closed-form solution. 
It is seen from Fig. \ref{CPUTime} that the designed BCD-ASO algorithm has the ability to save two orders of magnitude in computational time in contrast to the BCD-SOCP algorithm and obtains a high-quality solution, which demonstrates the efficiency of the MM-based algorithm.


\begin{figure}[t]
	\vspace{0.2cm}	
	\begin{center}
		\includegraphics[scale=0.65]{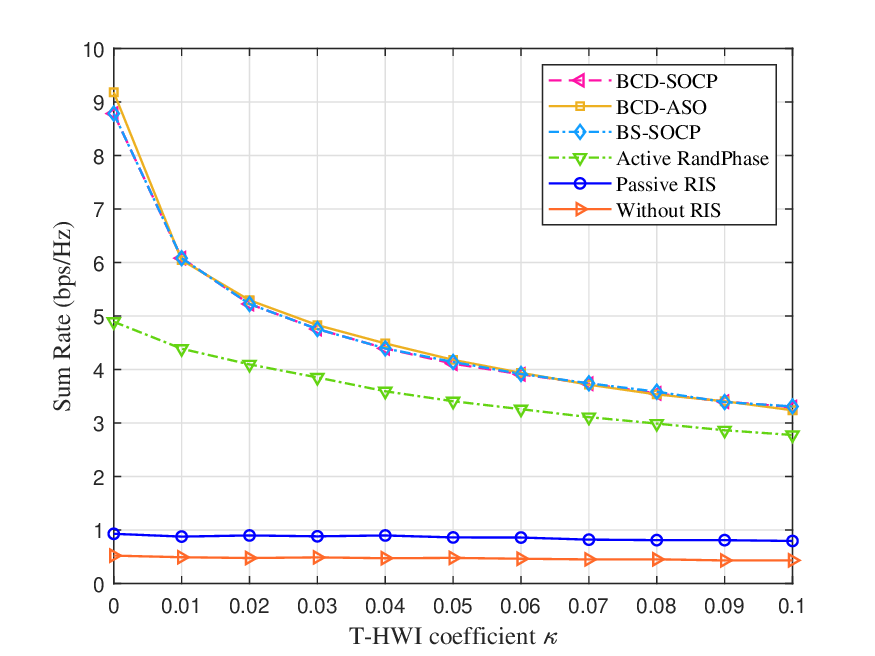}
		\caption{Sum rate versus the T-HWI coefficient $\kappa$.
		}
		\label{kappa}
	\end{center}
\end{figure}

\subsection{Impact of the Maximum Total Power Budget}

Fig. \ref{power} shows the sum rate performance versus the power budget $P$. In all schemes, the active/nearly passive RISs will not be activated until the total power budget meets the RIS circuit consumption. When the RIS is activated, the active RIS-empowered communication system outperforms the passive RIS-aided counterpart with the same total power budget and the same number of elements, owing to the fact that the elements of the active RIS possesses the ability of amplifying the incident signals. 
If the direct links are not too weak, deploying nearly passive RIS only brings a negligible performance gain, while the active RIS achieves a remarkable performance gain. For instance, when $P=40$ dBm and $\kappa = 0$, 
the passive RIS realizes $23\%$ gain, while the active RIS achieves noticeable $348\%$ gain. Besides, when the sum rates of the active RIS system and passive one are equal, the total power budget required by the former is much lower, which indicates that an active RIS is promising to save the total power consumption of the communication system.
Furthermore, the proposed algorithms achieve better system performance compared to the benchmark scheme proposed in \cite{zhang2021active}, thereby confirming the effectiveness of our proposed methods.
Moreover, the designed joint optimization algorithms obtain higher sum rates than the RandPhase scheme, verifying the advantages of the joint optimization. 
On the other hand, 
the negative impact of HWIs on the active RIS-assisted communications is more significant than that on the passive counterpart, which suggests the necessity of taking the HWIs into consideration in the active RIS design. 
Besides, with a high power budget $P$, the negative impact of HWIs on the performance of the active RIS-empowered communication is comparatively serious. The reason is that the distortion noise power is proportional to the signal power and that is amplified by the active RIS. 
Nevertheless, the performance gain brought by an active RIS is still greater than that achieved by the passive counterpart for the system with HWIs. 
It is also notable that the performance gain from increasing the total power budget decreases gradually, when the transceiver hardware and RIS are imperfect in an active RIS-empowered system.
In other words, a specific level of total power budget exists, beyond which little sum rate gain is obtained with the increase of $P$ in the active RIS system, when the HWIs are taken into account.

\subsection{Impact of the HWI coefficients}


\begin{figure}[t]
	\begin{center}
		\includegraphics[scale=0.65]{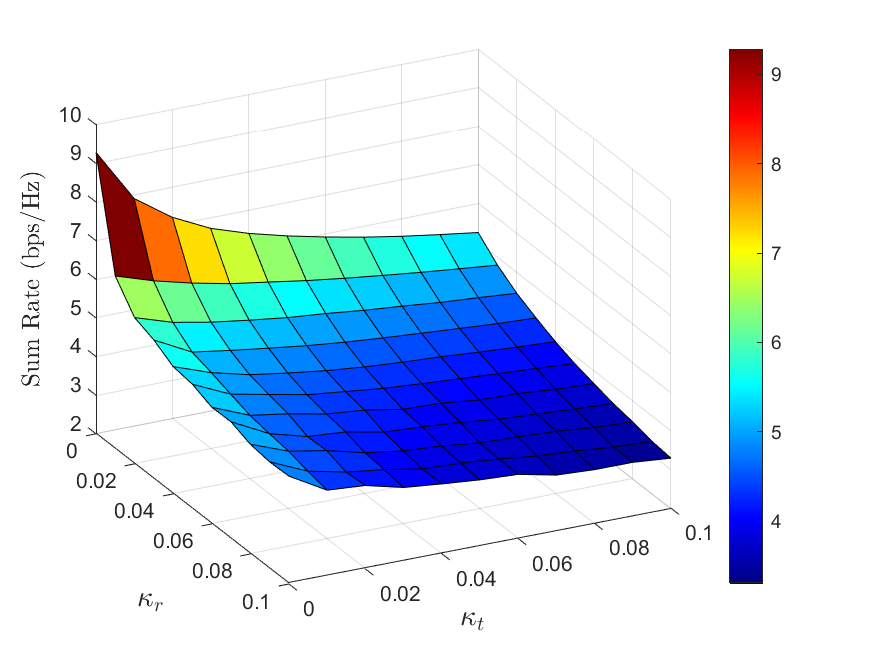}
		\caption{Sum rate versus $\kappa_t$ and $\kappa_r$ (3D).
		}
		\label{3D}
	\end{center}
\end{figure}

Fig. \ref{kappa} depicts the negative impact of the T-HWI coefficient $\kappa$ on the system performance. 
Specifically, the sum rates of different schemes deteriorate as $\kappa$ increase, due to the more prominent negative impact of HWIs. 
In the meantime, as $\kappa$ increases, the performance degradation of the active RIS system is more pronounced than that of the passive counterpart.
For instance, when $\kappa$ increase from 0 to 0.1, the sum rate of the active RIS system decreases by 65\% from 9.27 bps/Hz to 3.23 bps/Hz, while the sum rate of the nearly passive RIS system decreases by 14\%. Nevertheless, the performance gain brought by the active RIS is still significantly higher than that achieved by the passive one. In other words, a more noticeable performance gain is obtained by the active RIS in contrast to the nearly passive RIS for a relatively small $\kappa$.

\begin{figure}[t]
	\begin{center}
		\includegraphics[scale=0.65]{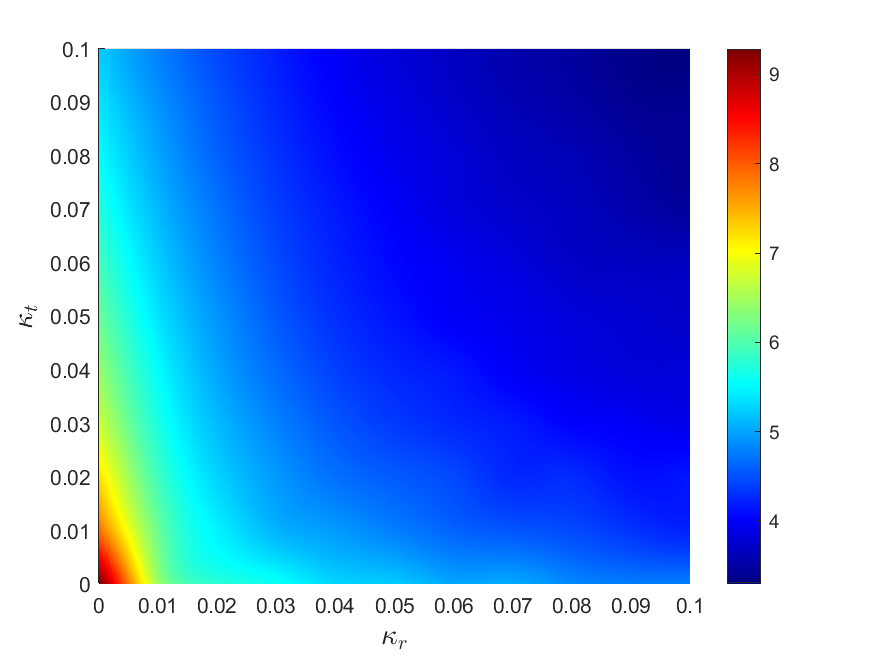}
		\caption{Sum rate versus $\kappa_t$ and $\kappa_r$ (2D).
		}
		\label{2D}
	\end{center}
\end{figure}

As illustrated in Fig. \ref{3D} and Fig. \ref{2D}, we evaluate the sum rate performance as a function of the transmitter HWI coefficient $\kappa_t$ and the receiver HWI coefficient $\kappa_r$. With the
increase of $\kappa_t$ and $\kappa_r$, the sum rate decreases as expected, which is consistent with the observation in Fig. \ref{kappa}. 
It can be observed that the level of hardware imperfection at the receiver poses a more severe negative impact on active RIS-empowered systems.
For example, in Fig. \ref{2D}, the system with $\left\{\kappa_r = 0.01, \kappa_t = 0\right\}$ is with the same sum rate performance as the system with $\left\{\kappa_r = 0, \kappa_t = 0.03\right\}$. 
The main reason is that the received distortion noise power is greater than the transmit distortion noise power, especially when the received signals are amplified by an active RIS.
%

\vspace{-0.3cm}	
\section{Conclusion}\label{Section 6}
In this work, we focused on the sum rate maximization problem in an active RIS-empowered communication system in the presence of both T-HWI and RIS-HWI. We jointly designed the transmit beamforming at the BS, along with the amplification factors and the phase shifts at the active RIS for maximizing the sum rate. To tackle the formulated non-convex optimization problem, the original problem  was decoupled into more tractable subproblems. Subsequently, we developed an SOCP-based algorithm, where the optimization variables are updated alternately. To further reduce the computational complexity, we exploited the MM-based Lagrange dual decomposition method and MM-based ASO approach for deriving the closed-form solution of each subproblem. Simulation results revealed that the active RIS is a promising technique to compensate for the multiplicative fading effect. Meanwhile, the results verified the effectiveness of the designed algorithms. Compared with the systems without RIS and with nearly passive RIS, the system with active RIS achieved higher sum rate under the same total power budget, which demonstrated the superiority of the active RIS. 
Meanwhile, our results indicated that the performance loss caused by the HWIs were more serious in the active RIS-empowered communication system than that in the passive one, which shows the necessity of taking the HWIs into consideration in communication system design. Nevertheless, the performance gain obtained by the active RIS is still more noticeable than that achieved by the nearly passive RIS. 

\bibliographystyle{IEEEtran}
\bibliography{IEEEabrv,Refer}
\begin{IEEEbiography}[{\includegraphics[width=1in,height=1.25in,clip,keepaspectratio]{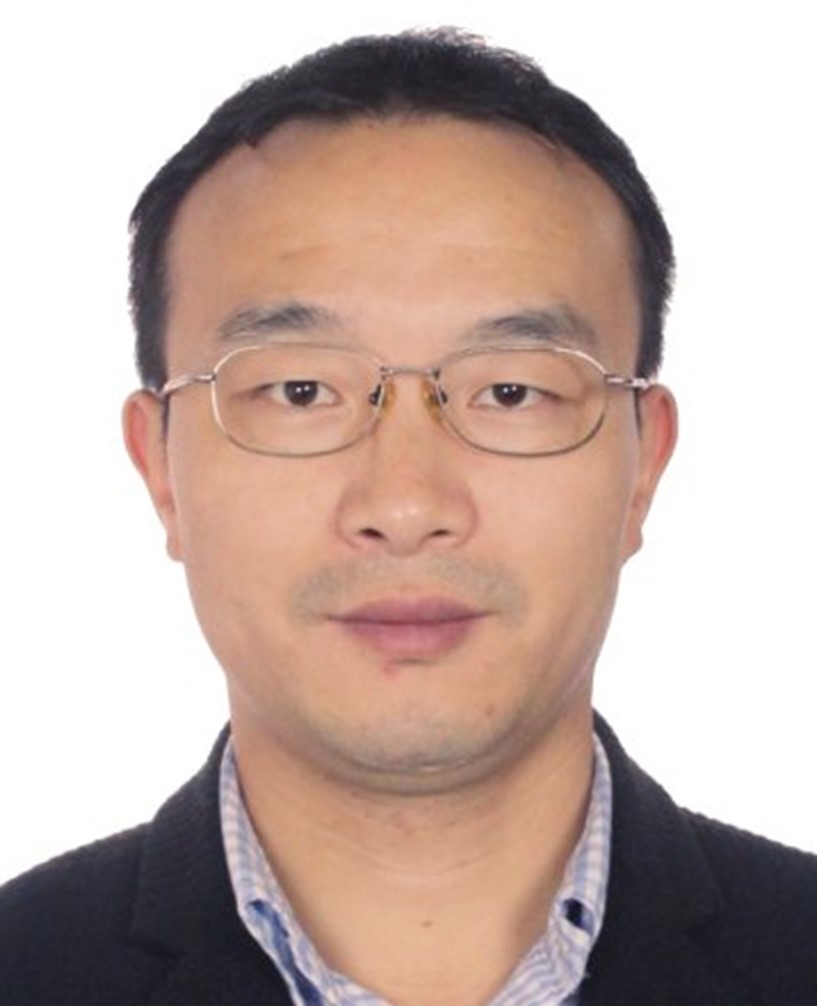}}]{Zhangjie Peng}
	received the B.S. degree from Southwest Jiaotong University, Chengdu, China, in 2004, and the M.S. and Ph.D. degrees from Southeast University, Southeast University, Nanjing, China, in 2007, and 2016, respectively, all in Communication and Information Engineering. He is currently an Associate Professor at the College of Information, Mechanical and Electrical Engineering, Shanghai Normal University, Shanghai 200234, China. 
	
	His research interests include reconfigurable intelligent surface (RIS), cooperative communications, information theory, physical layer security, and machine learning for wireless communications.
	
\end{IEEEbiography}

\begin{IEEEbiography}[{\includegraphics[width=1in,height=1.25in,clip,keepaspectratio]{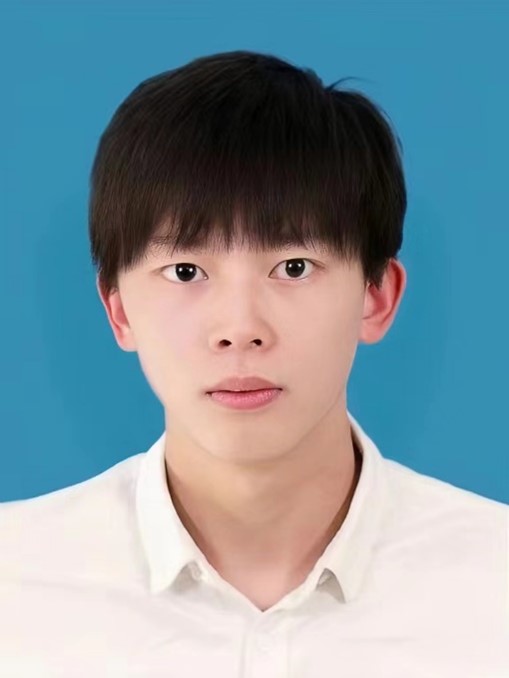}}]{Zhibo Zhang}
	received the B.S. degree from the College of Internet of Things Engineering, Hohai University, Changzhou, China, in 2020. He is currently pursuing the M.S. degree at the College of Information, Mechanical and Electrical Engineering, Shanghai Normal University, Shanghai, China. 
	
	His major research interests lie in the areas of communication and signal processing, including reconfigurable intelligent surface (RIS), physical layer security and machine learning for wireless communications. 
\end{IEEEbiography}

\begin{IEEEbiography}[{\includegraphics[width=1in,clip,keepaspectratio]{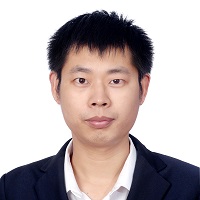}}]{Cunhua Pan} 
	is a full professor in Southeast University.  His research interests mainly include  reconfigurable intelligent surfaces (RIS),  AI for Wireless, and near field communications and sensing. He has published over 170 IEEE journal papers. His papers got over 13,000 Google Scholar citations with H-index of 59. He is  Clarivate Highly Cited researcher. He is/was an Editor of IEEE Transaction on Communications, IEEE Transactions on Vehicular Technology, IEEE Wireless Communication Letters, IEEE Communications Letters and IEEE ACCESS. He serves as the guest editor for IEEE Journal on Selected Areas in Communications on the special issue on xURLLC in 6G: Next Generation Ultra-Reliable and Low-Latency Communications. He also serves as a leading guest editor of IEEE Journal of Selected Topics in Signal Processing (JSTSP)  Special Issue on Advanced Signal Processing for Reconfigurable Intelligent Surface-aided 6G Networks, leading guest editor of IEEE Vehicular Technology Magazine on the special issue on Backscatter and Reconfigurable Intelligent Surface Empowered Wireless Communications in 6G, leading guest editor of IEEE Open Journal of Vehicular Technology on the special issue of Reconfigurable Intelligent Surface Empowered Wireless Communications in 6G and Beyond, and leading guest editor of IEEE IEEE Transactions on Green Communications and Networking Special Issue on Design of Green Near-Field Wireless Communication Networks. He received the  IEEE ComSoc Leonard G. Abraham Prize in 2022 and IEEE ComSoc Asia-Pacific Outstanding Young Researcher Award, 2022. 
	
\end{IEEEbiography}

\begin{IEEEbiography}[{\includegraphics[width=1in,clip,keepaspectratio]{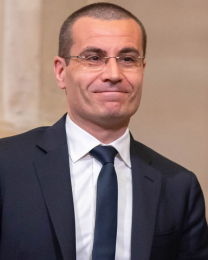}}]{Marco Di Renzo} 
	(Fellow, IEEE) received the Laurea (cum laude) and Ph.D. degrees in electrical
	engineering from the University of L’Aquila, Italy, in 2003 and 2007, respectively, and the Habilitation ã Diriger des Recherches (Doctor of Science) degree from Université Paris-Sud (currently Paris-Saclay University), Paris, France, in 2013. He was a Fulbright Fellow with The City University of New York, USA, a Nokia Foundation Visiting Professor in Finland, and a Royal Academy of Engineering Distinguished Visiting Fellow in U.K. He is currently a CNRS Research Director (Professor) and the Head of the Intelligent Physical Communications Group, Laboratory of Signals and Systems (L2S), CNRS and CentraleSupelec, Paris-Saclay University. He is also an Elected Member of the L2S Board Council and a member of the L2S Management Committee and the Admission and Evaluation Committee of the Ph.D. School on Information and Communication Technologies, Paris-Saclay University. He is a Founding Member and the Academic Vice Chair of the Industry Specification Group (ISG) on Reconfigurable Intelligent Surfaces (RIS), European Telecommunications Standards Institute (ETSI), where he served as the Rapporteur for the work item on communication models, channel models, and evaluation methodologies. He is a fellow of IET and AAIA, an Academician of AIIA, an Ordinary Member of the European Academy of Sciences and Arts and the Academia Europaea, and a Highly Cited Researcher. He is also serving as a Voting Member for the Fellow Evaluation Standing Committee and as the Director of Journals of the IEEE Communications Society. His recent research awards include the 2021 EURASIP Best Paper Award, the 2022 IEEE COMSOC Outstanding Paper Award, the 2022 Michel Monpetit Prize conferred by the French Academy of Sciences, the 2023 EURASIP Best Paper Award, the 2023 IEEE ICC Best Paper Award, the 2023 IEEE COMSOC Fred W. Ellersick Prize, the 2023 IEEE COMSOC Heinrich Hertz Award, the 2023 IEEE VTS James Evans Avant Garde Award, and the 2023 IEEE COMSOC Technical Recognition Award from the Signal Processing and Computing for Communications Technical Committee. He holds the 2023 France-Nokia Chair of Excellence in ICT. He served as the Editor-in-Chief for IEEE COMMUNICATIONS LETTERS from 2019 to 2023. He is serving on the Advisory Board.

\end{IEEEbiography}

\begin{IEEEbiography}[{\includegraphics[width=1in,clip,keepaspectratio]{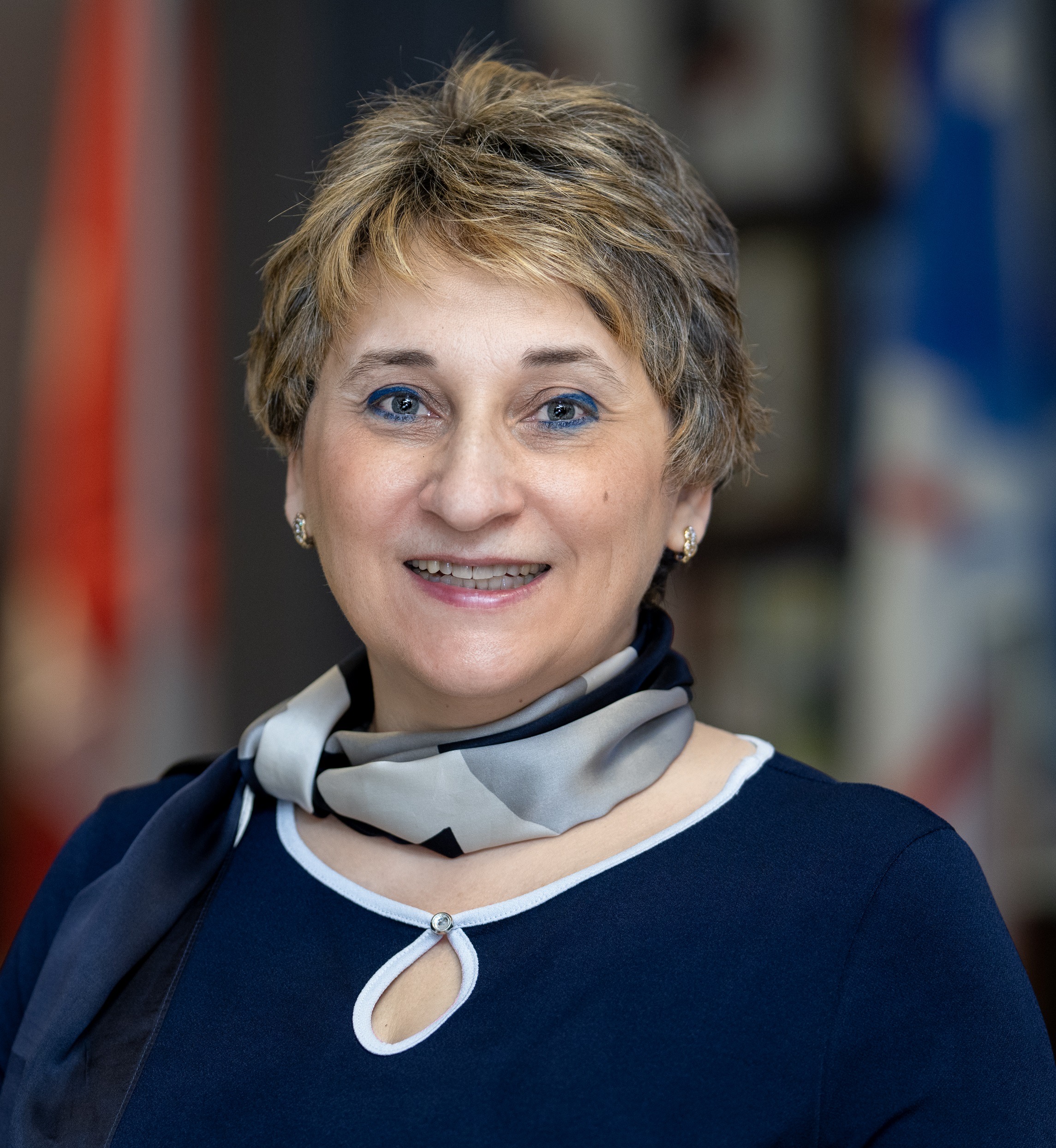}}]{Octavia A. Dobre} 
	 (Fellow, IEEE) is a Professor and Canada Research Chair Tier 1 with Memorial University, Canada. She was a Visiting Professor with Massachusetts Institute of Technology, USA and Université de Bretagne Occidentale, France.
	
	Her research interests encompass wireless communication and networking technologies, as well as optical and underwater communications. She has (co-)authored over 500 refereed papers in these areas.
	
	Dr. Dobre serves as the VP Publications of the IEEE Communications Society. She was the inaugural Editor-in-Chief (EiC) of the IEEE Open Journal of the Communications Society and the EiC of the IEEE Communications Letters.
	
	Dr. Dobre was a Fulbright Scholar, Royal Society Scholar, and Distinguished Lecturer of the IEEE Communications Society. She obtained Best Paper Awards at various conferences, including IEEE ICC, IEEE Globecom, IEEE WCNC, and IEEE PIMRC. Dr. Dobre is an elected member of the European Academy of Sciences and Arts, a Fellow of the Engineering Institute of Canada, and a Fellow of the Canadian Academy of Engineering.
	
\end{IEEEbiography}

\begin{IEEEbiography}[{\includegraphics[width=1in,clip,keepaspectratio]{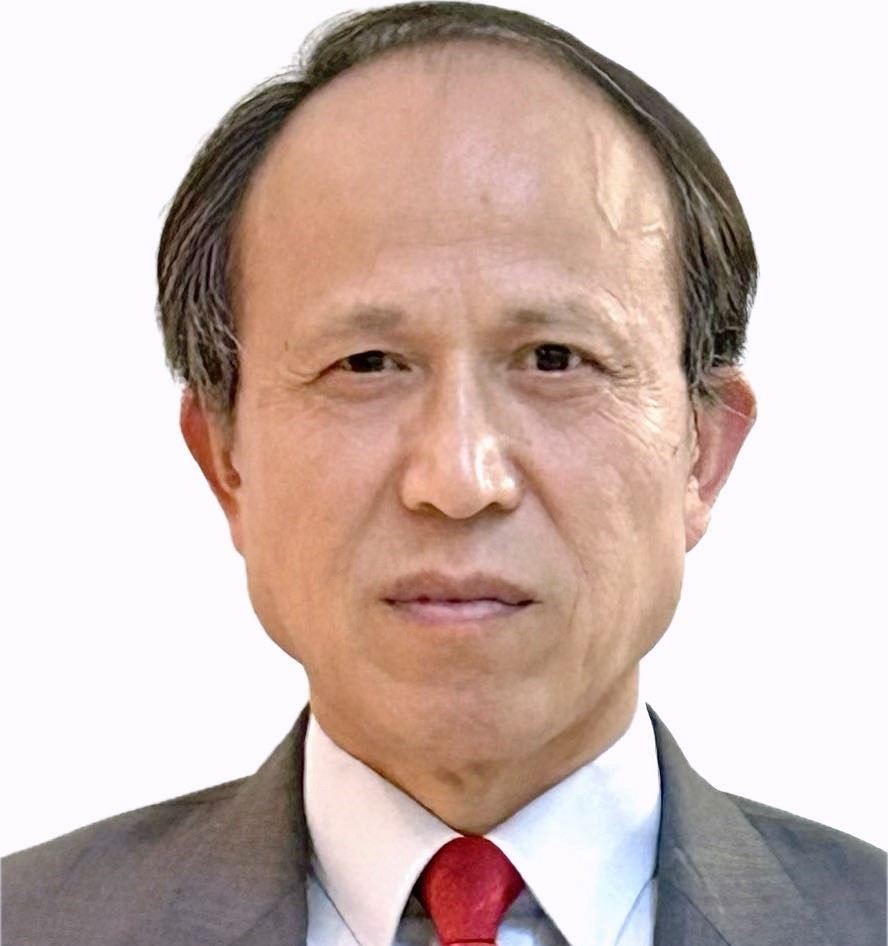}}]{Jiangzhou Wang} 
	(Fellow, IEEE) is a Professor with the University of Kent, U.K. He has published more than 500 papers and five books. His research focuses on mobile communications. He was a recipient of the 2022 IEEE Communications Society Leonard G. Abraham Prize. He was the Technical Program Chair of the 2019 IEEE International Conference on Communications (ICC2019), Shanghai, Executive Chair of the IEEE ICC2015, London, and Technical Program Chair of the IEEE WCNC2013. He is/was the editor of a number of international journals, including IEEE Transactions on Communications from 1998 to 2013. Professor Wang is a Foreign Member of the Chinese Academy of Engineering (CAE), a Fellow of the Royal Academy of Engineering (RAEng), U.K., Fellow of the IEEE, and Fellow of the IET. 
	
\end{IEEEbiography}

\end{document}